\documentclass[twocolumn]{aastex62}

\usepackage{verbatim}
\usepackage[usenames, dvipsnames]{}

\newcommand{\TESS}{\emph{TESS}}

\newcommand{\bjdtdb}{\ensuremath{\rm {BJD_{TDB}}}}

\newcommand{\kms}{km\,s$^{-1}$}

\newcommand{\logrphk}{\ensuremath{\log R^\prime_{\rm HK}}}

\newcommand{\mearth}{\ensuremath{M_\earth}}

\newcommand{\msun}{\ensuremath{M_\sun}}
\newcommand{\Msun}{\ensuremath{M_\sun}}
\newcommand{\ms}{\ensuremath{\rm m\,s^{-1}}}

\newcommand{\rearth}{\ensuremath{R_\earth}}

\newcommand{\rsun}{\ensuremath{R_\sun}}

\newcommand{\Teff}{$T_{\rm eff}$}
\newcommand{\teff}{$T_{\rm eff}$}
\newcommand{\jwst}{\emph{JWST}}

\newcommand{\thisstar}{TOI-1075}
\newcommand{\thisplanetb}{TOI-1075~b}
 
\newcommand{\plradunc}{1.791$^{+0.116}_{-0.081}$~\rearth}

\newcommand{\plmassunc}{9.95$^{+1.36}_{-1.30}$~\mearth}
 
\newcommand{\plrhounc}{$9.32^{+2.05}_{-1.85}$~g\,cm$^{-3}$}

\shorttitle{A Dense Hot Super-Earth around TOI-1075}
\shortauthors{Essack et al. 2022}

\accepted{October 19, 2022} 
\submitjournal{The Astronomical Journal}

\begin{document}

\title{TOI-1075~b: A Dense, Massive, Ultra-Short Period Hot Super-Earth Straddling the Radius Gap}

\author[0000-0002-2482-0180]{Zahra~Essack}
\affiliation{Department of Earth, Atmospheric and Planetary Sciences, Massachusetts Institute of Technology, Cambridge, MA 02139, USA}
\affiliation{Kavli Institute for Astrophysics and Space Research, Massachusetts Institute of Technology, Cambridge, MA 02139, USA}

\author[0000-0002-1836-3120]{Avi~Shporer}
\affiliation{Department of Physics and Kavli Institute for Astrophysics and Space Research, Massachusetts Institute of Technology, Cambridge, MA 02139, USA}

\author[0000-0002-0040-6815]{Jennifer~A.~Burt}
\affiliation{Jet Propulsion Laboratory, California Institute of Technology, 4800 Oak Grove Drive, Pasadena, CA 91109, USA}

\author[0000-0002-6892-6948]{Sara~Seager}
\affiliation{Department of Physics and Kavli Institute for Astrophysics and Space Research, Massachusetts Institute of Technology, Cambridge, MA 02139, USA}
\affiliation{Department of Earth, Atmospheric and Planetary Sciences, Massachusetts Institute of Technology, Cambridge, MA 02139, USA}
\affiliation{Department of Aeronautics and Astronautics, MIT, 77 Massachusetts Avenue, Cambridge, MA 02139, USA}

\author[0000-0001-6294-4523]{Saverio~Cambioni}
\affiliation{Department of Earth, Atmospheric and Planetary Sciences, Massachusetts Institute of Technology, Cambridge, MA 02139, USA}

\author[0000-0003-0525-9647]{Zifan~Lin}
\affiliation{Department of Earth, Atmospheric and Planetary Sciences, Massachusetts Institute of Technology, Cambridge, MA 02139, USA}

\author[0000-0001-6588-9574]{Karen~A.~Collins} 
\affiliation{Center for Astrophysics ${\rm \mid}$ Harvard {\rm \&} Smithsonian, 60 Garden Street, Cambridge, MA 02138, USA}

\author[0000-0003-2008-1488]{Eric~E.~Mamajek}
\affiliation{Jet Propulsion Laboratory, California Institute of Technology, 4800 Oak Grove Drive, Pasadena, CA 91109, USA}

\author[0000-0002-3481-9052]{Keivan~G.~Stassun} 
\affiliation{Department of Physics and Astronomy, Vanderbilt University, Nashville, TN 37235, USA}
\affiliation{Department of Physics, Fisk University, Nashville, TN 37208, USA}


\author[0000-0003-2058-6662]{George~R.~Ricker}
\affiliation{Department of Physics and Kavli Institute for Astrophysics and Space Research, Massachusetts Institute of Technology, Cambridge, MA 02139, USA}

\author[0000-0001-6763-6562]{Roland~Vanderspek}
\affiliation{Department of Physics and Kavli Institute for Astrophysics and Space Research, Massachusetts Institute of Technology, Cambridge, MA 02139, USA}

\author[0000-0001-9911-7388]{David~W.~Latham}
\affiliation{Center for Astrophysics ${\rm \mid}$ Harvard {\rm \&} Smithsonian, 60 Garden Street, Cambridge, MA 02138, USA}

\author[0000-0002-4265-047X]{Joshua~N.~Winn}
\affiliation{Department of Astrophysical Sciences, Princeton University, 4 Ivy Lane, Princeton, NJ 08544, USA}

\author[0000-0002-4715-9460]{Jon~M.~Jenkins}
\affiliation{NASA Ames Research Center, Moffett Field, CA 94035, USA}


\author[0000-0003-1305-3761]{R.~Paul~Butler}
\affiliation{Earth \& Planets Laboratory, Carnegie Institution for Science, 5241 Broad Branch Road, NW, Washington, DC 20015, USA}

\author[0000-0002-9003-484X]{David~Charbonneau}
\affiliation{Center for Astrophysics ${\rm \mid}$ Harvard {\rm \&} Smithsonian, 60 Garden Street, Cambridge, MA 02138, USA}

\author[0000-0003-2781-3207]{Kevin~I.~Collins}
\affiliation{George Mason University, 4400 University Drive, Fairfax, VA, 22030 USA}


\author[0000-0002-5226-787X]{Jeffrey~D.~Crane}
\affiliation{The Observatories of the Carnegie Institution for Science, 813 Santa Barbara Street, Pasadena, CA 91101, USA}

\author[0000-0002-4503-9705]{Tianjun~Gan}
\affil{Department of Astronomy, Tsinghua University, Beijing 100084, China}

\author{Coel~Hellier}
\affiliation{Astrophysics Group, Keele University, Staffordshire, ST5 5BG, UK}

\author[0000-0002-2532-2853]{Steve~B.~Howell}
\affiliation{NASA Ames Research Center, Moffett Field, CA 94035, USA}

\author{Jonathan~Irwin}
\affiliation{Institute of Astronomy, University of Cambridge, Madingley Road, Cambridge, CB3 0HA, UK}


\author[0000-0003-3654-1602]{Andrew~W.~Mann}
\affiliation{Department of Physics and Astronomy, The University of North Carolina at Chapel Hill, Chapel Hill, NC 27599-3255, USA}


\author[0000-0003-1102-1520]{Ali~Ramadhan}
\affiliation{Department of Earth, Atmospheric and Planetary Sciences, Massachusetts Institute of Technology, Cambridge, MA 02139, USA}

\author{Stephen~A.~Shectman}
\affiliation{The Observatories of the Carnegie Institution for Science, 813 Santa Barbara Street, Pasadena, CA 91101, USA}

\author{Johanna~K.~Teske}
\affiliation{Earth \& Planets Laboratory, Carnegie Institution for Science, 5241 Broad Branch Road, NW, Washington, DC 20015, USA}

\author[0000-0001-7961-3907]{Samuel~W.~Yee}
\affiliation{Department of Astrophysical Sciences, Princeton University, 4 Ivy Lane, Princeton, NJ 08544, USA}




\author[0000-0002-4510-2268]{Ismael~Mireles}
\affiliation{Department of Physics and Astronomy, University of New Mexico, 210 Yale Blvd NE, Albuquerque, NM 87106, USA}


\author[0000-0003-1309-2904]{Elisa~V.~Quintana}
\affiliation{NASA Goddard Space Flight Center, 8800 Greenbelt Road, Greenbelt, MD 20771, USA}

\author[0000-0002-1949-4720]{Peter~Tenenbaum}
\affiliation{NASA Ames Research Center, Moffett Field, CA 94035, USA}
\affiliation{SETI Institute, Mountain View, CA 94043, USA}

\author[0000-0002-5286-0251]{Guillermo~Torres}
\affiliation{Center for Astrophysics ${\rm \mid}$ Harvard {\rm \&} Smithsonian, 60 Garden Street, Cambridge, MA 02138, USA}

\author[0000-0001-9800-6248]{Elise~Furlan}
\affiliation{NASA Exoplanet Science Institute, Caltech/IPAC, Mail Code 100-22, 1200 E. California Blvd., Pasadena, CA 91125, USA}

\correspondingauthor{Zahra Essack}
\email{zessack@mit.edu}

\begin{abstract}
Populating the exoplanet mass-radius diagram in order to identify the underlying relationship that governs planet composition is driving an interdisciplinary effort within the exoplanet community. The discovery of hot super-Earths – a high temperature, short-period subset of the super-Earth planet population – has presented many unresolved questions concerning the formation, evolution, and composition of rocky planets. We report the discovery of a transiting, ultra-short period hot super-Earth orbiting \object{TOI-1075} (\object{TIC 351601843}), a nearby ($d$ = 61.4 pc) late K-/early M-dwarf star, using data from the Transiting Exoplanet Survey Satellite (\textit{TESS}). The newly discovered planet has a radius of \plradunc, and an orbital period of 0.605 days (14.5 hours). We precisely measure the planet mass to be \plmassunc\ using radial velocity measurements obtained with the Planet Finder Spectrograph (PFS), mounted on the Magellan II telescope. Our radial velocity data also show a long-term trend, suggesting an additional planet in the system. While \thisplanetb\ is expected to have a substantial H/He atmosphere given its size relative to the radius gap, its high density (\plrhounc) is likely inconsistent with this possibility. We explore \thisplanetb's location relative to the M-dwarf radius valley, evaluate the planet's prospects for atmospheric characterization, and discuss potential planet formation mechanisms. Studying the \thisstar\ system in the broader context of ultra-short period planetary systems is necessary for testing planet formation and evolution theories, density enhancing mechanisms, and for future atmospheric and surface characterization studies via emission spectroscopy with \jwst.
\end{abstract}


\section{Introduction} \label{sec:intro}

Hot super-Earths are a subset of the super-Earth planet population (1~R$_{\oplus }$ $<$ R$_{p}$ $<$ 2~R$_{\oplus }$), with short orbital periods (P $<$ 10 days), and surface temperatures high enough to melt silicate rock (T $>$ 800 K) due to strong irradiation by their host stars. Hot super-Earths are compelling objects to study for the insights that they provide into atmospheric loss/retention, volatile cycling, the behaviors of materials at extreme temperatures, and Earth’s early history as a magma-ocean planet.

NASA\textquotesingle s \textit{Kepler} space telescope \citep{borucki2010kepler} transformed our understanding of exoplanets with discoveries of new planet classes and planetary systems. One of the \textit{Kepler} mission’s most revolutionary scientific results was that among the short-period planets it was sensitive to (P $<$ 100 days), the size of the most common planet in the galaxy is between the size of Earth and Neptune (1-4~R$_{\oplus }$), which has no Solar System analog \citep{batalha2014exploring}. This population of planets is subdivided into super-Earths, 1~R$_{\oplus }$ $<$ R$_{p}$ $<$ 2~R$_{\oplus }$, and sub-Neptunes, 2~R$_{\oplus }$ $<$ R$_{p}$ $<$ 4~R$_{\oplus }$ \citep{fulton2017california}. The repurposing of the \textit{Kepler} mission into \textit{K2} \citep{howell2014k2} provided the opportunity to search for many more ultra-short period (USP) planets (P $<$ 1 day). The \textit{K2} Campaigns observed target fields in the ecliptic plane for 80 days at a time, making USP planets easily detectable within this observing window \citep{adams2016ultra, adams2017ultra, malavolta2018ultra, adams2021ultra}.

There are many unresolved theories regarding the atmospheres of hot super-Earths, including whether they exist (e.g. \citealp{kreidberg2019absence}), what they are composed of (e.g. \citealp{schaefer2009chemistry, ito2015theoretical, mansfield2019identifying}), and how they evolve. The “radius gap” -- a local minimum in the planet radius distribution at 1.75~R$_\earth$ for planets orbiting Sun-like stars, and with P $<$ 100 days \citep{fulton2017california} -- is theorized to separate predominantly rocky planets from planets with a substantial H/He atmosphere \citep{owen2017evaporation}. The location of the radius gap is dependent on the host star type, and shifts to a smaller radius as the stellar radius decreases, as seen for planets around M-dwarfs \citep{zeng2017gap, cloutier2020evolution}. \cite{rogers2015most} found that statistically planets with R$_{p} >$ 1.6~R$_\earth$ have a volatile-rich envelope. 
A variety of compositions have been determined for rocky planets below the radius gap, including Earth-like compositions (e.g. \citealp{pepe2013earth, dressing2015mass}) and high density compositions akin to Mercury (e.g. \citealp{santerne2018earth, lam2021gj, silva2022hd}). 

To explain the existence of the radius gap/valley, multiple theoretical models have been suggested. These include: photoevaporation -- atmospheric loss driven by stellar irradiation \citep{lopez2012thermal, chen2016evolutionary, owen2017evaporation}; core-powered mass loss -- atmospheric loss driven by the cooling of the planetary core after formation, resulting in the escape of the outer layers of the atmosphere \citep{ginzburg2018core,gupta2019sculpting,gupta2020signatures}; and gas-poor formation -- the formation of distinct rocky and non-rocky planet populations, where the rocky planet population is a result of delayed gas accretion within the protoplanetary disk until a point where the gas in the disk has almost fully dissipated \citep{lee2014make,lee2016breeding,lopez2018formation,cloutier2020evolution}. 

While planets discovered by \textit{Kepler} have well-constrained radii measurements, the vast majority lack corresponding mass measurements because the planets orbit faint stars, making detailed follow-up investigations difficult. NASA\textquotesingle s Transiting Exoplanet Survey Satellite (\TESS; \citealp{ricker2014transiting}) mission, the successor of \textit{Kepler}, is an all-sky survey of bright, nearby stars, with a minimum observing baseline of $\sim27$ days. \TESS\ has identified hundreds of short-period super-Earth planet candidates amenable to follow-up observations with radial velocity (RV) instruments to determine their masses since it began operations in 2018 \citep{guerrero2021tess}. Here we present the discovery and confirmation of \thisplanetb, an ultra-short period super-Earth around \thisstar\ ($V = 12.75$~mag) monitored by \TESS, with a planetary radius located slightly above the radius gap. Obtaining precise radii and masses for the small, close-in, \TESS\ planet candidates that span the radius valley is crucial for elucidating the atmospheric composition and evolution of hot super-Earths via further spectroscopic characterization, and for furthering our understanding of planetary compositions by studying planetary system architectures and formation histories.\\

This paper is structured as follows. In Section \ref{sec:stellardata}, we describe the properties of the host star \thisstar. In Section \ref{sec:planetdata}, we describe the time-series photometry and RV data sets we obtained for the \thisstar\ system. In Section \ref{sec:juliet_jointfit}, we describe our data analysis, including a global model fit, and derive properties for the planetary system. In Section \ref{sec:discussion}, we discuss the new star-planet system, including atmospheric characterization prospects and a review of potential formation mechanisms for \thisplanetb, and finally we provide our conclusions in Section \ref{sec:conclusions}.


\section{Stellar Data \& Characterization \label{sec:stellardata}}

\subsection{Astrometry \& Photometry}

Stellar astrometry and visible and infrared photometry for \thisstar\ (TIC~351601843, 2MASS~J20395334-6526579, APASS~31990797, Gaia~DR3~6426188308031756288, UCAC4~123-179251) are compiled in Table \ref{tab:star}. 
The positions, proper motions, parallax, radial velocity, and Gaia photometry are from Gaia DR3 \citep{prusti2016gaia, GaiaDR3}.
We convert the astrometry to Galactic velocities following 
\cite{ESA1997}\footnote{$U$ towards Galactic center, $V$ in direction of Galactic spin, and $W$ towards North Galactic Pole \citep{ESA1997}.}.
Photometry is reported from APASS Data Release 10 \citep{Henden2016}\footnote{https://www.aavso.org/apass}, the \TESS\ Input Catalog (TIC8; \citealp{Stassun2019}), 2MASS \citep{Cutri2003}, and WISE \citep{Cutri2012}.
From comparison of the star's colors ($B$-$V$ = 1.42, $V$-$K_s$ = 3.58, $G$-$K_s$ = 2.93, $V$-$J$  = 2.76, $G_{Bp}$-$G_{Rp}$ = 1.84) and absolute magnitude ($M_V$ = 8.76, $M_G$ = 8.10, $M_{Ks}$ = 5.17) with typical parameters for stars of various spectral types, \thisstar's photometry appears to be consistent with that of a typical main sequence star intermediate between K9V and M0V types \citep{Pecaut2013}.

\begin{table*}[tb]
    \caption{Astrometry \& Photometry for \thisstar}
    \centering
    \begin{tabular}{lcc}
    \hline
    Parameter & Value & Source \\
     \hline
    Designations & TIC 351601843 & \citet{Stassun2019}\\
    RA (ICRS, J2000) & 20:39:53.082 & Gaia DR3\\
    Dec (ICRS, J2000) & -65:26:58.95 & Gaia DR3\\
    $\mu$ RA (mas yr$^{-1}$) & -99.8399 $\pm$ 0.0081 & Gaia DR3\\
    $\mu$ Dec (mas yr$^{-1}$)& -60.016 $\pm$ 0.013 & Gaia DR3\\
    Parallax (mas) & 16.2816 $\pm$ 0.0132 & Gaia DR3\\
    Distance (pc) & $61.43^{+0.18}_{-0.67}$ & Gaia DR3\\
    $v_R$ (\kms) &  31.07 $\pm$ 0.30 & Gaia DR3\\
    Spectral Type & K9V/M0V & \citet{Pecaut2013}\\
    \hline
     $B$ (mag) & 14.108 $\pm$ 0.028 & APASS/DR10\\
     $V$ (mag) & 12.751 $\pm$ 0.077 & APASS/DR10\\
     $g'$ (mag) & 13.423 $\pm$ 0.027 & APASS/DR10\\
     $r'$ (mag) & 12.181 $\pm$ 0.088 & APASS/DR10\\
     $i'$ (mag) & 11.504 $\pm$ 0.143 & APASS/DR10\\
     {\it TESS} (mag) & 10.2565 $\pm$ 0.0074 & TICv8\\
      $G$ (mag) & 12.0447 $\pm$ 0.0028 & Gaia DR3\\
      $G_{BP}$ (mag) & 12.9442 $\pm$ 0.0028 & Gaia DR3\\
      $G_{RP}$ (mag) & 11.1069 $\pm$ 0.0038 & Gaia DR3\\
      $J$ (mag) & 9.935 $\pm$ 0.023 & 2MASS\\
      $H$ (mag) & 9.292 $\pm$ 0.026 & 2MASS\\
      $K_{s}$ (mag) & 9.115 $\pm$ 0.023 & 2MASS\\
      $W_{1}$ (mag) & 9.001 $\pm$ 0.025 & WISE\\
      $W_{2}$ (mag) & 9.001 $\pm$ 0.021 & WISE\\
      $W_{3}$ (mag) & 8.915 $\pm$ 0.024 & WISE\\
      $W_{4}$ (mag) & 8.806 $\pm$ 0.315 & WISE\\
\hline
    \label{tab:star} \end{tabular}
\end{table*}

\subsection{Spectral Energy Distribution \label{sec:SED}}

\begin{figure}[htb!]
    \centering
    \includegraphics[trim=70 80 80 80,clip,angle=90,width=\linewidth]{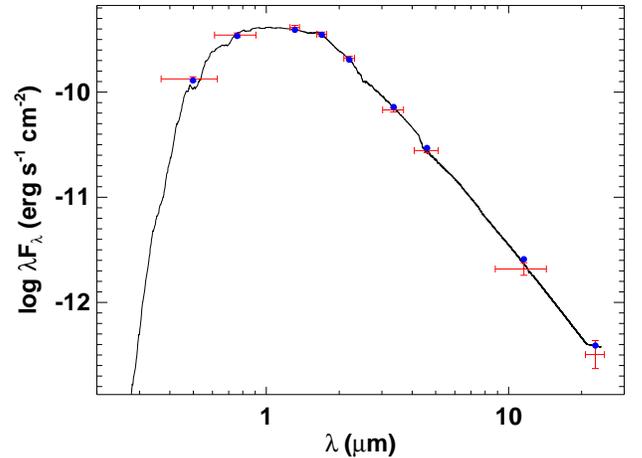}
\caption{Spectral energy distribution of \thisstar. Red symbols represent the observed photometric measurements, where the horizontal bars represent the effective width of the passband. Blue symbols are the model fluxes from the best-fit NextGen atmosphere model (black).  \label{fig:sed}}
\end{figure}

As an independent determination of the basic stellar parameters, we performed an analysis of the broadband spectral energy distribution (SED) of the star together with the Gaia DR3 parallax \citep[with no systematic offset applied; see, e.g.,][]{StassunTorres:2021}, in order to determine an empirical measurement of the stellar radius and mass following the procedures described in \citet{Stassun:2016,Stassun:2017,Stassun:2018}. We pulled the $JHK_S$ magnitudes from {\it 2MASS}, the W1--W4 magnitudes from {\it WISE}, and the $G$, $G_{\rm BP}$ and $G_{\rm RP}$ magnitudes from Gaia. Together, the available photometry spans the stellar SED over the wavelength range 0.4--20~$\mu$m (see Figure~\ref{fig:sed}).  

We performed a fit using NExtGen stellar atmosphere models, with the free parameters being the effective temperature ($T_{\rm eff}$) and metallicity ([Fe/H]). The remaining free parameter is the extinction $A_V$, which we fixed at zero due to the star's proximity\footnote{The
STILISM 3D reddening maps from \citet{Lallement2018} estimate the reddening towards \thisstar\ to be E(B-V) = $0.004\pm0.016$, i.e. negligible.}.
The resulting fit (Figure~\ref{fig:sed}) has a reduced $\chi^2$ of 1.4, with best fit $T_{\rm eff} = 3875 \pm 75$~K and [Fe/H] = $-0.5 \pm 0.5$. Integrating the model SED gives the observed bolometric flux, $F_{\rm bol} = 5.82 \pm 0.14 \times 10^{-10}$ erg~s$^{-1}$~cm$^{-2}$
($m_{\rm bol}$ = $11.590\pm0.026$ mag on IAU 2015 scale).
Adopting the Gaia DR3 parallax ($\varpi$ = $16.2816\pm0.0132$ mas), this leads to a bolometric luminosity of log($L_{\rm bol}$/$L_{\odot}$) = $-1.163\pm0.010$. 
Combining the luminosity with the derived $T_{\rm eff}$ provides an estimate of the stellar radius of $R_\star = 0.581\pm0.024$~$R_\odot$. 
In addition, we can estimate the stellar mass from the empirical $M_K$-based relations of \citet{Mann:2019}, which give $M_\star = 0.604 \pm 0.030$~M$_\odot$. Moreover, the radius and mass together imply a mean stellar density of $\rho_\star = 4.34 \pm 0.57$~g~cm$^{-3}$. 



\subsection{Stellar Mass and Radius from Empirical Relations}

While we adopt the host star parameters derived above (Section \ref{sec:SED}), based on the SED and NExtGen models, we estimated those parameters using the empirical relations of \cite{Mann:2019} and \cite{Boyajian:2012}, for comparison. 
We used the Gaia DR3 distance to derive the absolute $K$ band magnitude from the observed 2MASS magnitude, resulting in $M_K = 5.173 \pm 0.023$ mag. 

Next, we used the empirical relation between stellar mass and $M_K$ provided by \citet[][see their Table 6 and Equation 2]{Mann:2019}. Assuming a conservative uncertainty of $5\%$ this resulted in $M_\star = 0.571 \pm 0.029\ \msun$. We note that the empirical relations provided by \cite{Mann:2019} cover a mass range that reaches about 0.75 \msun\ and a $K$ band magnitude up to about 4 mag, so the TOI 1075 host star is well within that range. For comparison, we calculated the stellar mass with the empirical relation of \citet[see their Table 1 and Equation 10]{Mann:2015}, resulting in $0.604 \pm 0.030\ \msun$, which is 5\% from the above estimate.

To estimate the stellar radius we used the mass derived above and the radius-mass empirical relation derived by \citet[][their Equation 10]{Boyajian:2012}, resulting in $0.541 \pm 0.027\ \rsun$. For comparison, using the empirical relation between radius and $M_K$ of \citet[][see their Table 1]{Mann:2015}, results in $R_\star = 0.580 \pm 0.029\ \rsun$, which is 7\% or 1.3$\sigma$ larger than the above estimate. \\

For the subsequent analyses in this paper, we adopt the stellar parameters derived from the SED analysis following \citet{Stassun:2016,Stassun:2017,Stassun:2018}, namely $R_\star = 0.581\pm0.024 R_\odot$, $M_\star = 0.604 \pm 0.030$~M$_\odot$, and $T_{\rm eff} = 3875 \pm 75$~K.


\subsection{Speckle Observations}

If a star hosting a planet candidate has a closely bound stellar companion (or companions), the companion can create a false-positive exoplanet detection if it is a stellar eclipsing binary (EB). Additionally, flux from these companion source(s) can lead to an underestimated planetary radius if not accounted for in the transit model \citep{ciardi2015understanding}. To search for close-in bound companions unresolved in our other follow-up observations, we obtained high resolution speckle imaging observations. 

\begin{figure}[!ht]
    \centering
    \includegraphics[width=\linewidth]{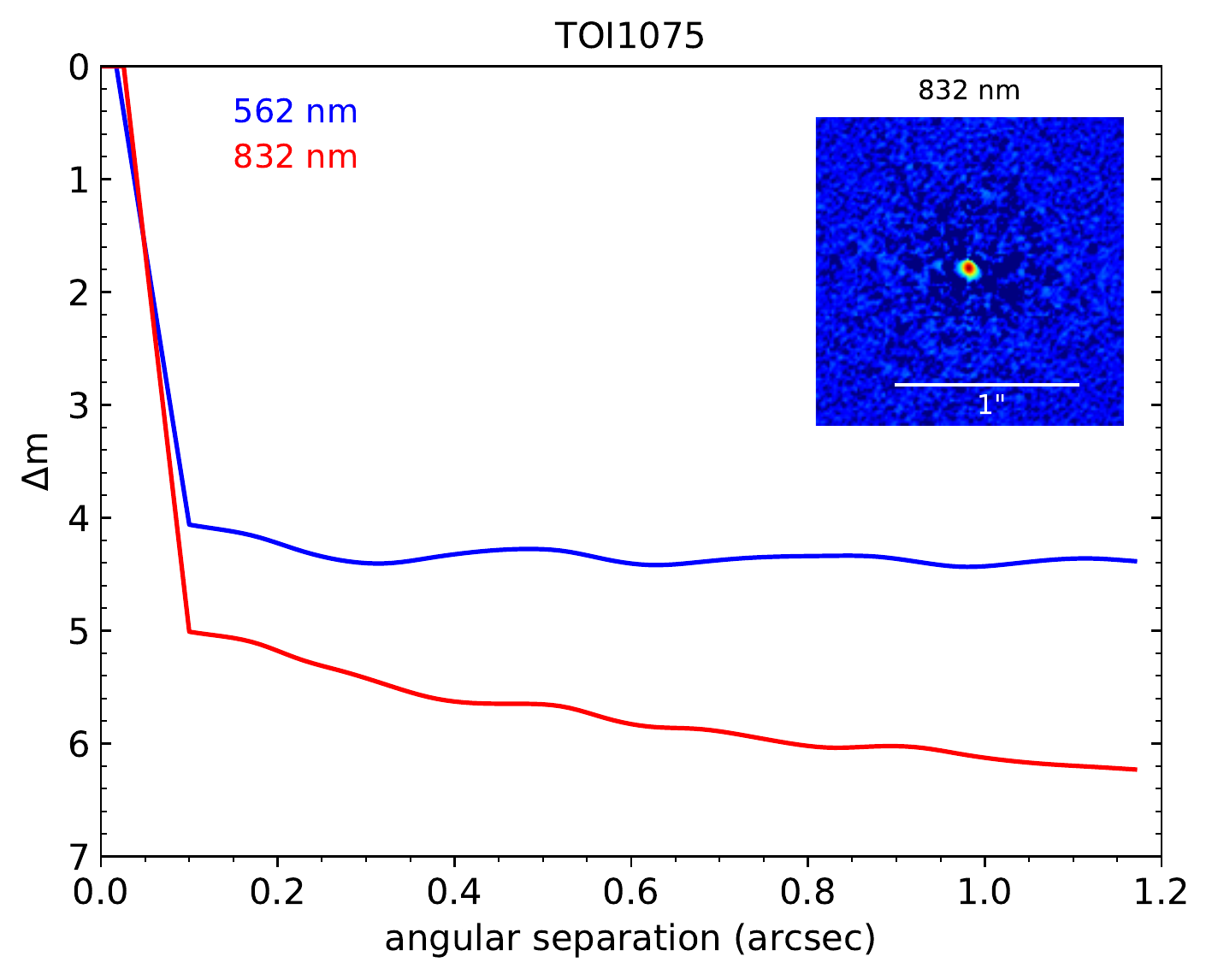}
\caption{Gemini South Zorro contrast curves for 562 nm (blue line) and 832 nm (red line) speckle observations, and 832 nm image (inset) of \thisstar. No visual companions are detected anywhere in Zorro's field of view. \label{fig:speckle}}
\end{figure}

\thisstar\ was observed on 2019 September 12 UT using the Zorro speckle instrument on Gemini South \citep{scott2021twin}. Zorro provides simultaneous speckle imaging in two bands (562 $\pm$ 54 nm and 832 $\pm$ 40 nm) with output data products including a reconstructed image and robust contrast limits on companion detections \citep{howell2011speckle,howell2016speckle}. Figure \ref{fig:speckle} shows the 5-sigma limiting contrast curves for the Zorro observations in both 562 nm (blue line) and 832 nm (red line), and the 832 nm reconstructed speckle image. We find that \thisstar\ is a single star with no companion brighter than $\delta$m = 5--6 mag at 832 nm from about 0.1" out to 1.2". At the distance to \thisstar\ ($d=61.4$~pc), these angular limits correspond to spatial separations of 6 to 74 AU.

\subsection{Stellar Kinematics and Population}

\citet{GaiaDR3} provides the most accurate position and proper motion for \thisstar\ (=Gaia~DR3~6426188308031756288) and \citet{BailerJones2021} converts the Gaia DR3 trigonmetric parallax ($\varpi = 16.282\pm0.013$ mas) into a geometric distance of $d = 61.43^{+0.18}_{-0.67}$ pc. 
\citet{GaiaDR2} lists a median radial velocity of $v_r$ = $31.07 \pm 0.30$ \kms\, averaged over 24 epochs. 
Using formulae from \citet{ESA1997}, we convert the Gaia DR3 astrometry and radial velocity into Galactic barycentric velocities: 
$U, V, W$ = 33.49, -31.58, 7.49 ($\pm$0.22, 0.30, 0.20) \kms\,\footnote{The velocities are within 0.01 \kms\, of that reported in Gaia GCNS catalog \citet{GCNS2021}.}.
The star's 3D velocity is not near any of the 29 nearby young stellar groups tracked by \citet{Gagne2018}, and the BANYAN $\Sigma$ tool\footnote{http://www.exoplanetes.umontreal.ca/banyan/banyansigma.php}
returns membership probabilities of zero ($<$0.1\%).
Using the formulae and parameters from \citet{Bensby2014}, we estimate Galactic population kinematic membership probabilities for TOI-1075 of P(thin disk) = 98.2\%, P(thick disk) = 1.8\%, P(halo) = $4.2\times10^{-3}$\%, P(Hercules) = $1.3\times10^{-2}$\%, i.e. TOI-1075 is
$\sim$56$\times$ more likely to be a thin disk star than a thick disk star based on its velocity alone. %
The oldest thin disk stars are approximately $\sim$8-9 Gyr \citep[e.g.][]{Kilic2017,Fuhrmann2017,Fantin2019,Tononi2019}.

We calculated the 3D separation between \thisstar\ and all the stars in the Gaia Catalogue of Nearby Stars \citep{GCNS2021} to search for any potential stellar companions. 
We find that \thisstar\ has no neighbors within 2\,pc, 
and Gaia~DR3~6429596764016919296 ($\Delta$ = 2.00\,pc; component of the tight binary 2MASS 20335172-6403200) is its nearest star. 

The systematic survey for common proper motion companions to stars within 100\,pc by \citet{Kervella2022} did not yield any matches of \thisstar\ with any {\it Hipparcos} stars. 
A query of the \citet{GaiaDR3} catalog within 2$^{\circ}$ ($\sim$2.0\,pc) of \thisstar\ searching among stars with parallaxes and proper motions within 25\%\, of the values for \thisstar\ yielded no plausible common proper motion companions. 
Thus, \thisstar\ appears to be a single star.\\


The space velocity of \thisstar\ may provide some additional clues about its age.
In the GCNS catalog of stars within 100\,pc, there are 153 stars whose $UVW$ velocities are within 10 \kms\, of \thisstar. 
Thirty eight of the 153 stars are lacking SIMBAD entries and have not been noted in the literature. 
Among the 115 with SIMBAD entries, 60 have fiducial spectral types in SIMBAD 
and none are hotter than the F5V star HD~43879 \citep[\teff\, = $6566\pm86$ K;][]{Casagrande2011}. 
A query of stellar parameter catalogs with mass estimates \citep{Schofield2019,Stassun2019,Paegert2021,Reiners2022}, 
shows that among the $d$ $<$ 100\,pc stars with velocities within 10\,\kms\, of \thisstar, there is a noticeable lack of stars more massive than 1.40\,\msun\footnote{
HIP~29888 (HD~43879; $1.38\pm0.23$ \Msun) in the TICv8.2 \citep{Paegert2021}, and 
HIP~29888, HIP~111971 (HD~214729), and HIP~70196 (HD~125346) in \citep{Reiners2020} - all with mass estimates of
1.40 \Msun\, - defining the upper mass envelope.}.
The list of GCNS stars with velocities within 10\,\kms\, of \thisstar\ was also queried through the compendium of chromospheric activity measurements (\logrphk) from \citet{BoroSaikia2018}.
Only 8 of the stars had \logrphk\, measurements, and only two had \logrphk\, $>$ -4.8 \citep[approximately corresponding
to the Sun on its very most active days;][]{Egeland2017}: the planet host star HD~128356 (HIP~71481; K2.5IV, \logrphk\, = -4.73) and 6~And (HD~218804, HIP~114430; F5V, \logrphk\, = -4.52). 
6~And is a 2.99$^{+0.48}_{-0.99}$ Gyr-old, fast-rotating \citep[$v$sin$i$ = 19\,\kms;][]{Schroeder2009} F5V star near the Kraft break, and so not unusually fast-rotating or young. 
The \logrphk\, value for HD~128356 (-4.73) appears to be spurious, however, as the assumed B-V color from Hipparcos (0.685) is based
on a single ground-based measurement \citep{Mermilliod1997} which is at odds with the star's spectral type 
\citep[K3V or K2.5IV;][]{Upgren1972,Gray2006} and the star's \Teff, for which the published estimates are in tight agreement
\citep[4932-4953\,K;][]{Sousa2018,Luck2018,Soto2018}. 
Adopting the B-V estimate for HD~128356 from the Tycho catalog ($1.04\pm0.02$ mag), which is more consistent with that for a K3 dwarf star, the median Mt.~Wilson S-value quoted by \citet[][S=0.214]{BoroSaikia2018} translates \citep[via formulae from ][]{Noyes1984} to a more benign chromospheric activity level of \logrphk\, = -5.06. 
Hence, the stars that have 3D velocities within 10\,\kms\, of \thisstar\ that are Sun-like (excluding the mid-F star 6~And) with Ca H \& K indices all have \logrphk\, $<$ -4.8, consistent with ages of $\gtrsim$3 Gyr \citep{Mamajek2008}.\\

We conclude that the lack of $>$1.40\,\msun\, stars with similar velocities to \thisstar\ suggests that stars with similar orbits are unlikely to be $\lesssim$2 Gyr, and the small number of stars with published chromospheric activity indices seem to tell a similar story (lacking in stars $\lesssim$3 Gyr).
It appears that stars younger than $\lesssim$2 Gyr have not yet scattered into the velocity space adjacent to the orbit of \thisstar, suggesting that the star is either a middle-aged or old thin disk star, likely with an age between 2-9 Gyr.

\subsection{Metallicity}\label{sec:metallicity}

\thisstar\ was spectrally characterized by the RAVE \citep[RAVE~J203953.3-652658,][]{Kordopatis2013,Kunder2017,Steinmetz2020}
and GALAH \citep{Buder2018,Buder2021} stellar spectroscopy surveys, from which wildly disparate metallicity estimates have been published, ranging from [M/H] = $-0.99\pm0.09$ \citep{Kunder2017} to [Fe/H] = $0.38\pm0.07$ \citep{Buder2021}.

An independent photometric estimate of the metallicity can be made based on the star's position on a color-magnitude diagram. Using the $V-K_s$ vs. $M_{Ks}$ calibration from \citet{Johnson2009} and \cite{Schlaufman2010}, we find that the color-mag position for \thisstar\ ($V-K_s$ = 3.584, $M_{Ks}$ = 5.176) is only 0.056 mag below the mean main sequence for late K- and M-dwarfs, translating to photometric metallicity estimates of [Fe/H] = -0.08 \citep{Johnson2009} and -0.21 \citep{Schlaufman2010}. A similar estimate can be done using 2MASS and Gaia photometry by interpolating the photometry and metallicities of nearby M-dwarfs in \citet{Mann:2015}. As with $V-K_s$ vs. $M_{Ks}$, the star sits slightly below the Solar-metallicity sequence in $M_G$ using $B_P-R_P$, $J-K$, or $G-K_S$. This interpolation method gave us a metallicity estimate of [Fe/H] = -0.10$\pm$0.12, consistent with the first estimate from \citet{Johnson2009}. 

We also analyzed the iodine-free PFS template spectrum using the publicly available code \texttt{SpecMatch-Emp} \citep{Yee2017}. This code matches a target spectrum with a library of observed spectra from stars with empirically determined stellar properties, and is particularly well-suited for the analysis of late-type stars. We recovered $T_{\rm eff} = 3824 \pm 70$~K, $R_\star = 0.57\pm0.06~R_\odot$, and $\mathrm{[Fe/H]} = -0.08 \pm 0.09$~dex. The \texttt{SpecMatch-Emp} recovered metallicity thus corroborates the photometric metallicity measurement.



\subsection{Stellar Variability}

\begin{figure}[!ht]
    \centering
    \includegraphics[width=\linewidth]{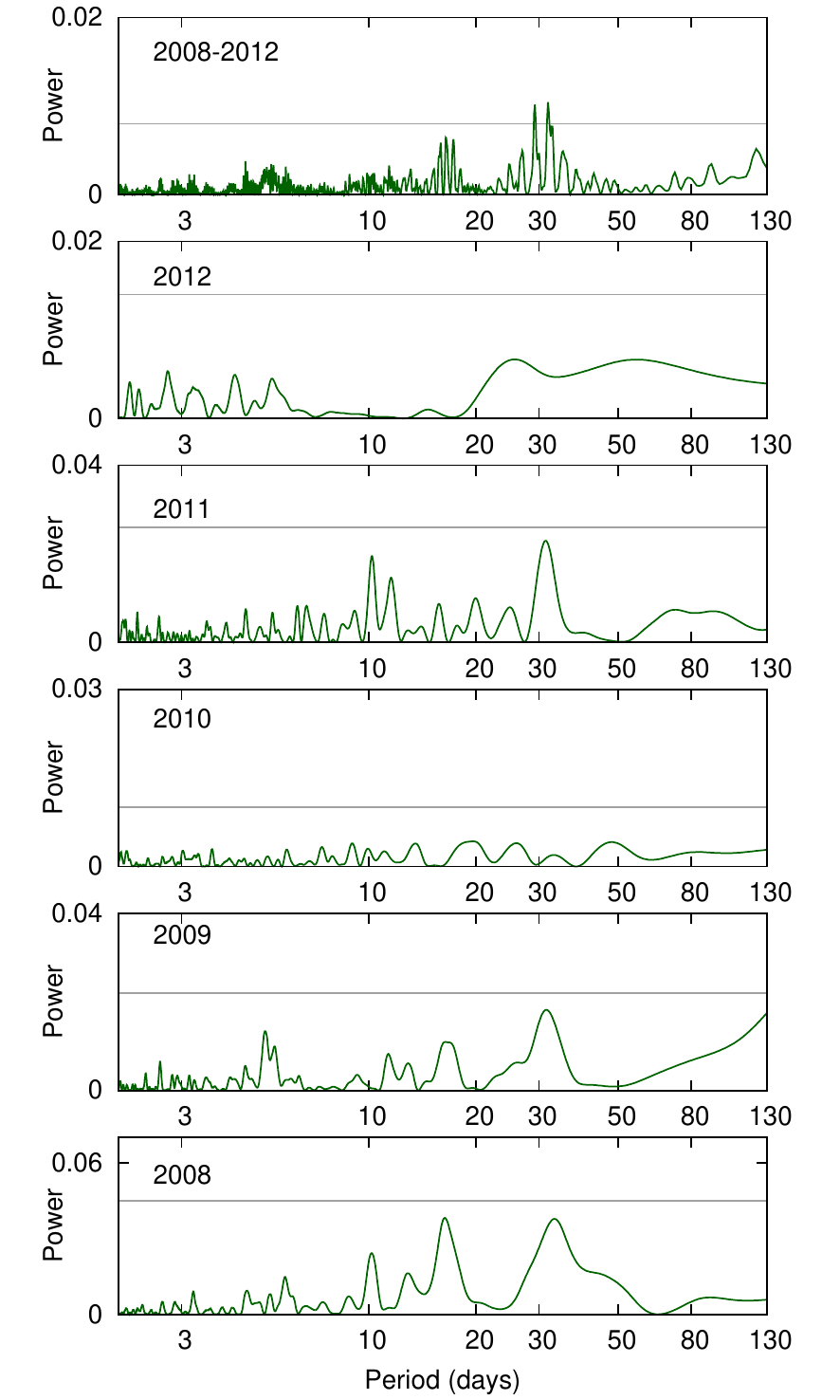}
\caption{Periodograms of each season of WASP data for \thisstar, showing the absence of any significant rotational modulation.  The horizontal lines mark the 95\%-confidence false-alarm level.  The top panel is the periodogram for the 5 years of data combined. \label{fig:wasp}}
\end{figure}

WASP-South, an array of 8 cameras composed of Canon 200-mm, f/1.8 lenses backed by 2k$\times$2k CCDs, was the Southern station of the WASP transit-search survey \citep{2006PASP..118.1407P}. The field of \thisstar\ was observed every year from 2008 to 2012, covering spans of 100 d to 180 d in each year. Within each night the cadence was typically 15-min, accumulating a total of 38\,000 data points.  \thisstar\ is the only bright star in the 48\arcsec\, photometric extraction aperture. We searched the WASP data for a rotational modulation using the methods described in \citet{2011PASP..123..547M} but we find no significant modulation for any period between 1 and 100 days. Within each season the 95\%-confidence upper limit on the amplitude is 3~mmag. Combining all the years of data results in an upper limit of 1.6~mmag. The periodograms of each season of WASP-South data showing no significant rotation modulation are shown in Figure \ref{fig:wasp}. The peaks near 30~days are compatible with the residual effects of moonlight propagating through the pipeline at a low level, and thus are unlikely to be caused by \thisstar.
The lack of any rotational modulation is unusual for a cool star.  \citet{2014ApJS..211...24M} report that 83\%\ of stars cooler than 4000~K in the Kepler field show a rotational modulation, with most having amplitudes in the range 3--10~mmag.  Hence \thisstar\ is among the least photometrically variable $\sim$\,20\%\ of stars of its spectral type. \thisstar's low levels of stellar variability support the finding that the star is at least 2~Gyr old.

\section{Exoplanet Detection \& Follow Up \label{sec:planetdata}}

\subsection{\TESS\ Time Series Photometry}

The \TESS\ primary mission surveyed the northern and southern ecliptic hemispheres in sectors measuring 24$^{\circ} \times$ 96$^{\circ}$, with near-continuous photometric coverage over $\sim$27~days. The TESS Primary Mission ran for 2 years (July~2018 -- July~2020), and consisted of 26~sectors. \TESS\ began its first extended mission in July 2020 which will end in September 2022 (when the second extended mission is scheduled to commence), and consists of 29~sectors. 
\thisstar\ (TIC~351601843, 2MASS~J20395334-6526579) was selected for transit detection observations by \TESS\ with 2-minute cadence as part of the Candidate Target List (CTL) -- a pre-selected target list prioritized for the detection of small planets \citep{stassun2018tess, Stassun2019}. 
\thisstar\ was observed by \TESS\ in Sector~13 from UT 2019 June 19 through 2019 July 18 during the primary mission, and again from UT 2020 July 4 through 2020 July 30 in Sector 27 during the first \TESS\ Extended Mission. The star fell on Camera~2 in both sectors.

The raw \TESS\ data for \thisstar\ were processed with the Science Processing Operations Center Pipeline (SPOC; \citealp{Jenkins2016}) -- which performs pixel calibration, light curve extraction, de-blending from nearby stars, and removal of common-mode systematic errors -- and are available at the Mikulski Archive for Space Telescopes (MAST) website\footnote{https://mast.stsci.edu}. The SPOC data include both simple aperture photometry (SAP) flux measurements \citep{Twicken2010, Morris2017} and presearch data conditioned simple aperture photometry (PDCSAP) flux measurements \citep{Smith2012, Stumpe2012, Stumpe2014}. The instrumental variations present in the SAP flux are removed in the PDCSAP flux data. 

\begin{figure*}[htb!]
    \centering
    \includegraphics[width=\textwidth,keepaspectratio]{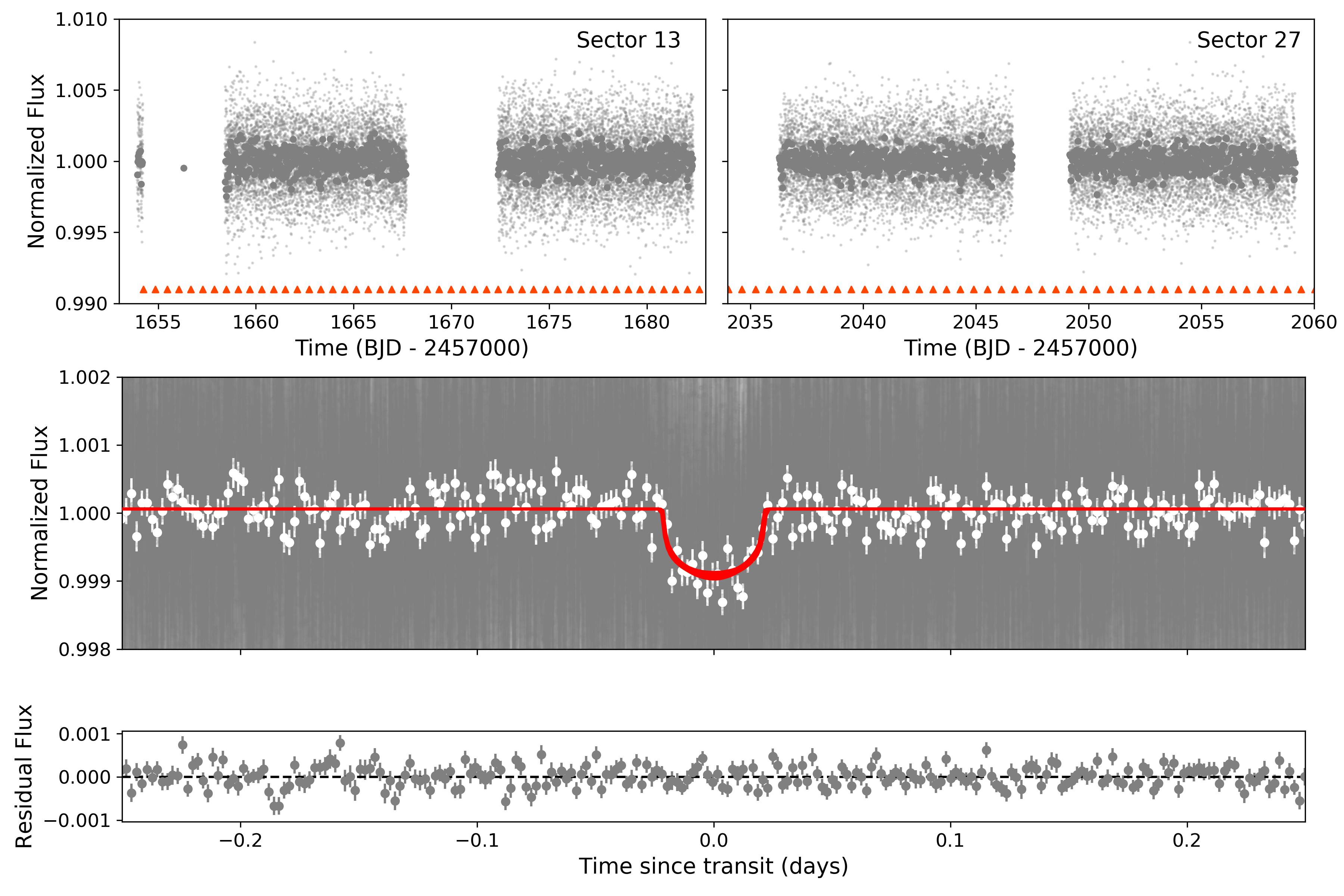}
\caption{\TESS\ light curves of \thisstar. \textit{Top:} Detrended, normalized and flattened PDCSAP flux measurements for Sector~13 (left) and Sector~27 (right). Lighter gray points are the \TESS\ 2-minute cadence flux measurements; darker points are the same data binned into 30-minute intervals. The transits of \thisplanetb\ are marked by orange triangles. \textit{Middle}: Phase-folded light curve on planet's orbital period (0.605~days), along with the best-fit transit model (red). Gray points are 2-minute cadence measurements; white points are the same data binned into 5-minute intervals. \textit{Bottom:} Residuals after the data have been subtracted from the best-fit model. \label{fig:tessdataphased}}
\end{figure*}

We further detrended the \TESS\ PDCSAP data by median normalizing, flattening, fitting a low order spline, and removing 3-$\sigma$ outliers from each sector's flux measurements separately, before stitching the light curves together. The detrended \TESS\ PDSCAP data for Sector~13 and Sector~27, as well as the phase-folded light curve are shown in Figure \ref{fig:tessdataphased}.

\subsubsection{\TESS\ Transit Detection}
The SPOC Transiting Planet Search (TPS; \citealp{jenkins2002impact, jenkins2010transiting}) pipeline searches for threshold crossing events (TCEs) in the PDCSAP light curve, applying an adaptive noise-compensating matched filter to account for stellar variability and residual observation noise. TCEs with a period of 0.605~days were detected independently in the SPOC transit search of the Sector~13 light curve\footnote{The gap at the beginning of the TESS Sector~13 data in Figure \ref{fig:tessdataphased} is a result of cadences being excluded from TPS due to the effects of rapidly changing scattered light and glints from the Earth and Moon.}, Sector~27 light curve, and multisector light curves from Sectors 13 and 27. 

In order to rule out false positives that can mimic the planetary transit signal, we evaluated the star's data validation reports (DVR; \citealp{twicken2018kepler, li2019kepler}), which are generated from the SPOC 2-minute cadence data. The multisector DVR shows no evidence of secondary eclipses, odd/even transit depth inconsistencies, or correlations between the depth of the transit and the size of the aperture used to extract the light curve -- which would indicate that the transit signal originated from a nearby eclipsing binary. Additionally, the location of the transit source as shown in the DVR is consistent with the position of the target star -- the difference image centroiding test located the source of the transits due to \thisplanetb\ to within $1.64\pm4.69$” of \thisstar, which complements the speckle imaging observations. Upon passing these vetting checks, the transit signal was assigned the identifier TOI-1075.01 and announced by the \TESS\ TOI team. \citep{guerrero2021tess}.


\subsection{Ground-based Time-Series Photometry}

After it was alerted as a TOI, we acquired ground-based time-series follow-up photometry of \thisstar\ during future times of transit predicted by the \TESS\ data. We used the {\tt TESS Transit Finder}, which is a customized version of the {\tt Tapir} software package \citep{Jensen:2013}, to schedule our transit observations.

\begin{figure*}[htb!]
    \centering
    \includegraphics[width=\textwidth,keepaspectratio]{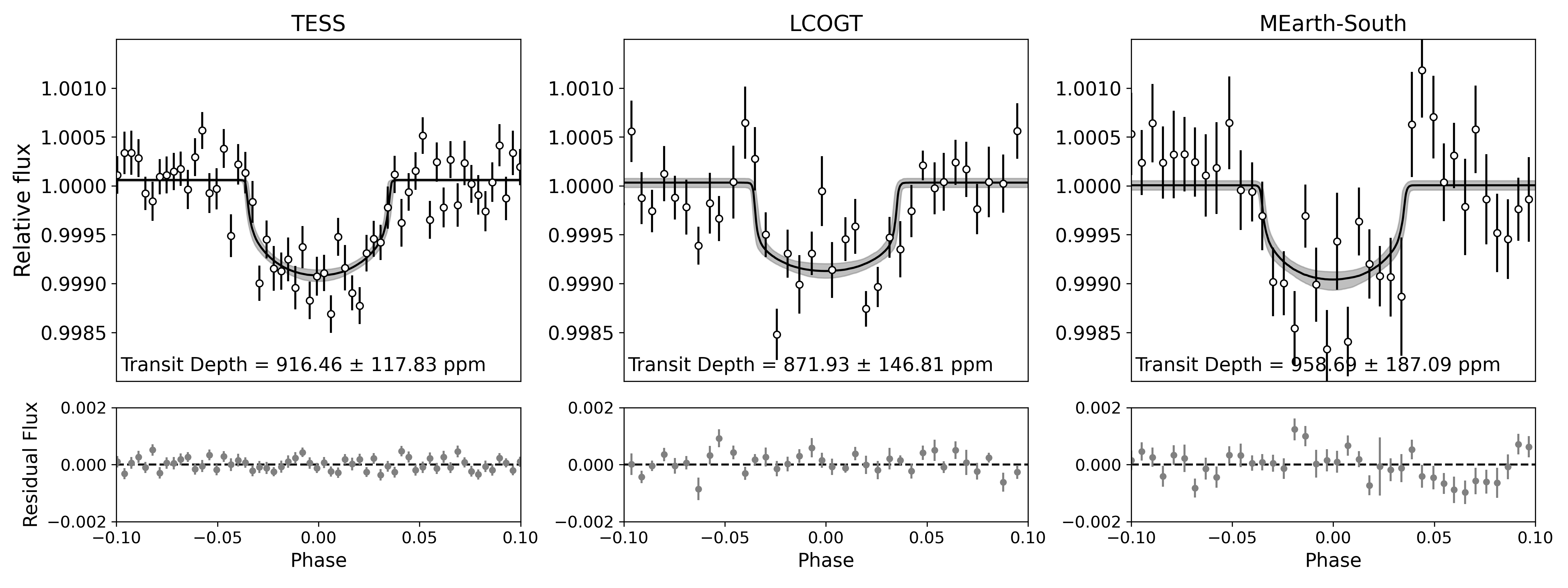}
\caption{Results of the \texttt{juliet} joint fit to the \TESS\ photometry, and ground-based LCOGT and MEarth-South photometry. \textit{Top:} Transit observations of the respective instruments phase-folded to the period of \thisplanetb. The black curve is the best-fit \texttt{juliet} transit model, and the 68\% confidence interval is represented by the gray shaded region. The binned data points with errorbars are shown for clarity (white circles). The \TESS\ data are binned in 2-minute intervals, and the LCOGT and MEarth-South data are binned in 5-minute intervals. \textit{Bottom:} Residuals after the data have been subtracted from the best-fit model. \label{fig:transitsjointfit}}
\end{figure*}

\subsubsection{LCOGT 1\,m Observations}

We observed five full transits of \thisstar\ in Pan-STARRS $z$-short band from the Las Cumbres Observatory Global Telescope \citep[LCOGT;][]{Brown:2013} 1.0\,m network on UTC 2019 August 25, 2019 August 26, 2019 September 23, 2019 September 24, and 2019 September 26 (Figure \ref{fig:transitsjointfit}). The first, third and fifth observations were conducted from the South African Astronomical Observatory (SAAO) node, and the second and fourth observations from the Siding Spring Observatory(SSO) node. The 1\,m telescopes are equipped with $4096\times4096$ SINISTRO cameras having an image scale of $0\farcs389$ per pixel, resulting in a $26\arcmin\times26\arcmin$ field of view. The images were calibrated by the standard LCOGT {\tt BANZAI} pipeline \citep{McCully:2018}. Photometric data were extracted using {\tt AstroImageJ} \citep{Collins:2017} and circular photometric apertures with radii in the range $6\farcs2$ to $7\farcs8$. The target star apertures exclude flux from all known nearby Gaia~DR3 and TESS Input Catalog stars. We detect the event on-target in all five data sets, which are included in the joint model of the system in this work.

\subsubsection{MEarth-South 0.4m Observations}

We observed two full transits of \thisplanetb\ using the MEarth-South telescope array \citep{nutzman2008design, irwin2015mearth} at Cerro Tololo Inter-American Observatory (CTIO), Chile on UT 2019 September 22 and 2019 September 28 (Figure \ref{fig:transitsjointfit}). Observations were gathered for approximately 5.5 hours centered on the predicted time of mid-transit. The data were reduced using the standard MEarth processing pipeline (e.g. \citealt{berta2012transit}) with a photometric extraction aperture of $r = 14$ pixels (11.8\arcsec). Twelve light curves were observed across 6 telescopes, and were collected with an RG715 filter. All of the light curves contain meridian flips prior to the predicted time of ingress.  These were accounted for in the analysis of the light curves by allowing for separate magnitude zero points for each combination of telescope and side of the meridian to remove any residual flat fielding error. Some residuals in the out of transit baseline, likely due to color-dependent atmospheric extinction, were found so the final model also included a linear decorrelation against airmass.

\subsubsection{Previous Validation of \thisplanetb}
Additionally, planet candidate TOI-1075.01 was statistically validated as a planet using \TESS\ and ground-based photometry in \citet{giacalone2022validation}. TOI-1075.01 was vetted with \texttt{DAVE} \citep{kostov2019discovery} which uses centroid offset analyses to identify evidence of false positives due to contamination from nearby stars, and with \texttt{TRICERATOPS} \citep{giacalone2020triceratops,giacalone2021vetting}, which calculates the Bayesian probability that the candidate is an astrophysical false positive. TOI-1075.01 showed no strong indicators of being a false positive in the aforementioned analysis, and was then validated as \thisplanetb\ \citep{giacalone2022validation}.


\subsection{Time Series Radial Velocities}

We collected 18 precision radial velocity epochs of \thisstar\ using the Planet Finder Spectrograph \citep[PFS;][]{crane2006,crane2008,crane2010} on the 6.5~m Magellan~II (Clay) telescope at Las Campanas Observatory in Chile. PFS is a slit-fed spectrograph that is wavelength calibrated using an iodine cell, and covers the wavelength range $391-734$~nm, though only the 500-620~nm range is used when measuring RV shifts. All PFS spectra are reduced and RVs extracted using a custom {\texttt{IDL}} pipeline based on \citet{butler1996} that regularly delivers sub-1 m~s$^{-1}$ precision. Each of the 18 PFS \thisstar\ iodine observations consist of 3$\times$20-minute exposures, and were mostly obtained on a night-by-night basis between UT 2021 May 22 and UT 2021 November 15, although in some cases two observations were taken on the same night but separated in time by at least 1-2 hours. A two-hour, iodine-free template observation was obtained on UT 2021 May 29. All observations were taken with the default 0.3$\arcsec$ slit but in 3$\times$3 binning mode, resulting in a resolving power of R$\sim$110,000. The resulting unbinned (20-minute integration time) RVs have typical precisions of 2.5--3.0 m~s$^{-1}$, and are listed in full in Table \ref{tab:RVs}.

\begin{figure}[htb!]
    \centering
    \includegraphics[width=\linewidth,keepaspectratio]{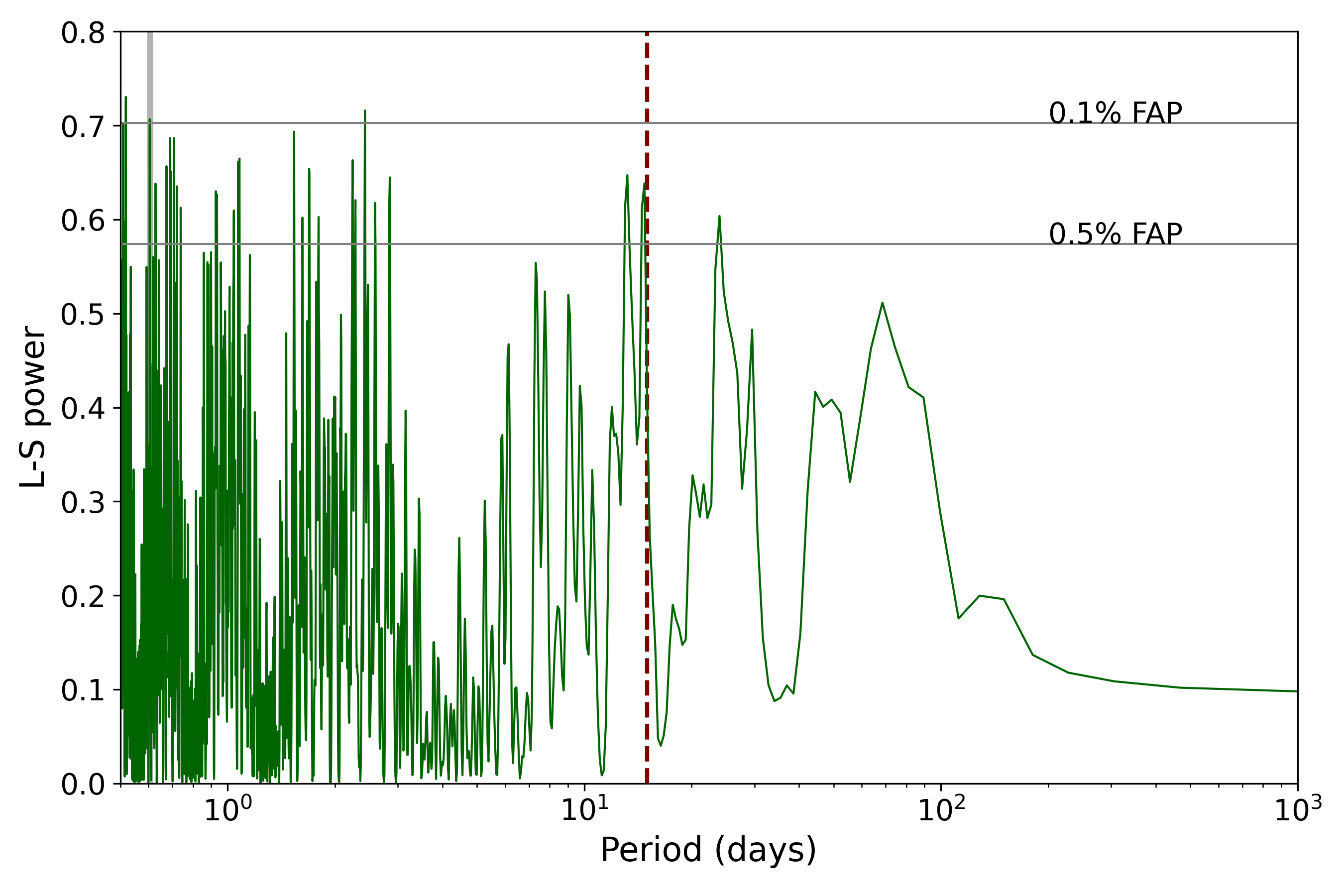}
\caption{Generalized Lomb–Scargle periodogram of the radial velocity measurements from PFS. There is a clear peak at the period of \thisplanetb\ (gray bar).  We identify a stellar activity signal at $\sim$15~days (red dashed line) which is well-separated from \thisplanetb's orbital period. \label{fig:PFSperiodogram}}
\end{figure}

A generalized Lomb–Scargle (GLS) periodogram of the PFS RV data shows a significant peak at 0.605~days (Figure \ref{fig:PFSperiodogram}), which matches the orbital period of the planet candidate determined from the \TESS\ data. An additional significant peak at $\sim$15~days likely corresponds to a stellar activity signal. The stellar activity signal is well-separated from the planet period, and there are no stellar periodicity signals more dominant than the planet signal, nor any significant stellar signals that appear around the planet period. Additionally, we modeled the suggested stellar activity signal using a Gaussian Process (GP) and find no evidence to incorporate it into our final model (see Section \ref{subsubsec:rvmodelcomp}). Therefore, incorporating GPs for non-white noise models, or adding another term to our RV model fit was deemed unnecessary. 


\begin{center}
\begin{longtable}{ccc}
\caption{PFS RV data of \thisstar}\\
\hline Date (BJD$_{\rm TDB}$) & RV ($\ms$) & $\sigma_{RV}$ ($\ms$)\\
\hline \vspace{2pt}
2459356.868 & -12.48 & 2.90 \\
2459356.884 & -5.77 & 3.04 \\
2459356.897 & -14.11 & 2.65 \\
2459357.856 & -25.83 & 3.00 \\
2459357.870 & -8.09 & 2.52 \\ 
2459357.885 & -18.11 & 3.38 \\
2459358.848 & -30.56 & 2.40 \\
2459358.862 & -26.80 & 2.18 \\
2459358.876 & -27.47 & 2.30 \\
2459452.564 & -11.29 & 2.34 \\
2459452.578 & -11.31 & 2.49 \\
2459452.593 & -6.06 & 2.57 \\
2459452.665 & -12.61 & 2.69 \\
2459452.679 & -5.05 & 2.56 \\
2459452.693 & -6.42 & 3.01 \\
2459470.617	& 10.02	& 3.18 \\
2459470.631 & -6.18	& 3.05 \\
2459470.645 & 1.55 & 3.65 \\
2459471.561 & 10.02 & 2.61 \\
2459471.576 & 14.14 & 2.58 \\
2459471.590 & 13.93 & 2.55 \\
2459471.659 & 15.38 & 2.71 \\
2459471.673 & 13.14 & 2.52 \\
2459471.687 & 6.22 & 2.58 \\
2459473.537 & 11.41 & 2.48 \\
2459473.550 & 17.32 & 2.52 \\
2459473.565 & 20.98 & 3.25 \\
2459473.647 & 6.82 & 3.02 \\
2459473.661 & 10.98 & 3.19 \\
2459473.675 & 14.55 & 3.39 \\
2459474.554 & -3.83 & 2.69 \\
2459474.568 & 4.77 & 2.62 \\
2459474.583 & -2.40 & 3.05 \\
2459474.616 & 6.56 & 3.00 \\
2459474.630 & 2.55 & 3.02 \\
2459475.526 & -5.85 & 3.07 \\
2459475.540 & -3.90 & 2.47 \\
2459475.554 & -6.89 & 2.60 \\
2459501.537 & 7.60 & 2.90 \\
2459501.551 & -2.56 & 2.80 \\
2459501.565 & 4.48 & 2.86 \\
2459504.528 & 0.00 & 2.47 \\
2459504.542 & -9.94 & 2.75 \\
2459504.556 & -12.43 & 2.91 \\
2459505.549 & 11.69 & 4.34 \\
2459505.563 & 7.41 & 3.99 \\
2459505.578 & 4.74 & 3.25 \\
2459506.518 & -19.93 & 2.41 \\
2459506.532 & -1.76 & 2.45 \\
2459506.546 & -14.89 & 2.33 \\
2459531.523 & 7.30 & 2.49 \\
2459531.537 & 10.63 & 2.55 \\
2459533.536 & 7.42 & 2.67 \\
2459533.546 & 11.35 & 4.27 \\
\hline
\label{tab:RVs}
\end{longtable}
\end{center}

\section{System Parameters from \texttt{\lowercase{juliet}}}
\label{sec:juliet_jointfit}

To obtain precise parameters of the \thisstar\ system, we performed a joint analysis of the \TESS, LCOGT, and MEarth-South photometry, and the PFS RV data using \texttt{juliet} \citep{espinoza2019juliet}. \texttt{juliet} is a fitting tool that uses nested sampling algorithms to efficiently sample a given parameter space, and allows for model comparison based on Bayesian evidences. \texttt{juliet} combines the publicly available packages for transits and RV modeling, \texttt{batman} \citep{kreidberg2015batman} and \texttt{radvel} \citep{fulton2018radvel}, respectively. We opted to implement \texttt{dynesty} \citep{speagle2020dynesty} as the nested sampling algorithm for our joint fitting, though a range of nested sampling algorithms are available to choose from.

\subsection{Transit Modeling}
For the transit modeling, \texttt{juliet} employs the \texttt{batman} package. We adopt a quadratic model to describe the limb darkening effect in the \TESS, LCOGT, and MEarth-South photometry, and parameterize it by employing the efficient, uninformative sampling scheme of \citet{kipping2013efficient}, and a quadratic law. We used a fixed dilution factor of 1 for all instruments, but considered free individual instrumental offsets. Instrumental jitter terms were taken into account and added in quadrature to the nominal instrumental errorbars.
We used uniform priors per the \citet{espinoza2018efficient} parameterization to explore the full physically plausible parameter space for the planet-to-star radius ratio, $p=R_{p}/R_{\star}$, and impact parameter, $b$.  
Additionally, we defined a log-uniform prior on the stellar density, and then recovered the scaled semi-major axis ($a/R_*$) using Kepler's third law.

\subsection{RV Modeling}
The model that we selected for our RV joint fit analysis was composed of a circular Keplerian orbit for the transiting planet (ultra-short period planets are expected to be tidally circularized), and an additional linear long-term trend to constrain the non-Keplarian long-period signal present in the PFS RV data, whose period is longer than the current observation baseline. We assumed uniform wide priors for the systemic velocity, jitter term, and RV semi-amplitude of the PFS RVs, as well as the linear long-term trend parametrized by an intercept $B$, and a slope $A$. We measured a radial velocity semi-amplitude of $K = 10.95^{+1.50}_{-1.43}~\ms$ for \thisplanetb.

\subsubsection{RV Model Comparison} \label{subsubsec:rvmodelcomp}

\texttt{juliet} searches for the global posterior maximum based on the evaluation of the Bayesian log-evidence ($\ln\mathcal{Z}$), allowing us to perform model comparisons given the differences in $\Delta\ln\mathcal{Z}$. We modeled the PFS RV data with and without fitting a linear long-term trend to the data. The model with the linear long-term trend had a log-evidence of $\ln\mathcal{Z}$ = $-$204.01 $\pm$ 0.38, and the model without the linear long-term trend (a circular Keplarian model only) had a log-evidence of $\ln\mathcal{Z}$ = $-$212.95 $\pm$ 0.28, resulting in a $\Delta\ln\mathcal{Z}$ = 8.94. We selected the model with the linear long-term trend component following criteria described in \citet{trotta2008bayes}, which considers a $\Delta\ln\mathcal{Z}$ $>$ 2 as weak evidence that one model is preferred over the other, and a $\Delta\ln\mathcal{Z}$ $>$ 5 as strong evidence that one model is significantly preferred over the other, hence the additional model parameters are necessary to account for the long-period signal in the PFS RV data. We also considered a model composed of a circular Keplerian orbit, and a long-term trend parametrized by an intercept $B$, a slope $A$, and a quadractic/curvature coefficient, $Q$. The change in log-evidence was $\Delta\ln\mathcal{Z}$ $<$ 2 between the long-term trend parametrization with and without the additional parameter, $Q$, confirming that the RV data are legitimately fitted by a circular Keplerian model and a linear long-term trend. As an additional test, we added a stellar signal to the RV model using a GP to determine the effect of the 15-day period signal identified in the PFS GLS periodogram (Figure \ref{fig:PFSperiodogram}). We compared the models with and without the additional stellar signal and found no evidence ($\Delta\ln\mathcal{Z}$ $<$ 2) that the final model required the addition of a stellar signal.

\begin{figure*}[htb!]
    \centering
    \includegraphics[width=\textwidth,keepaspectratio]{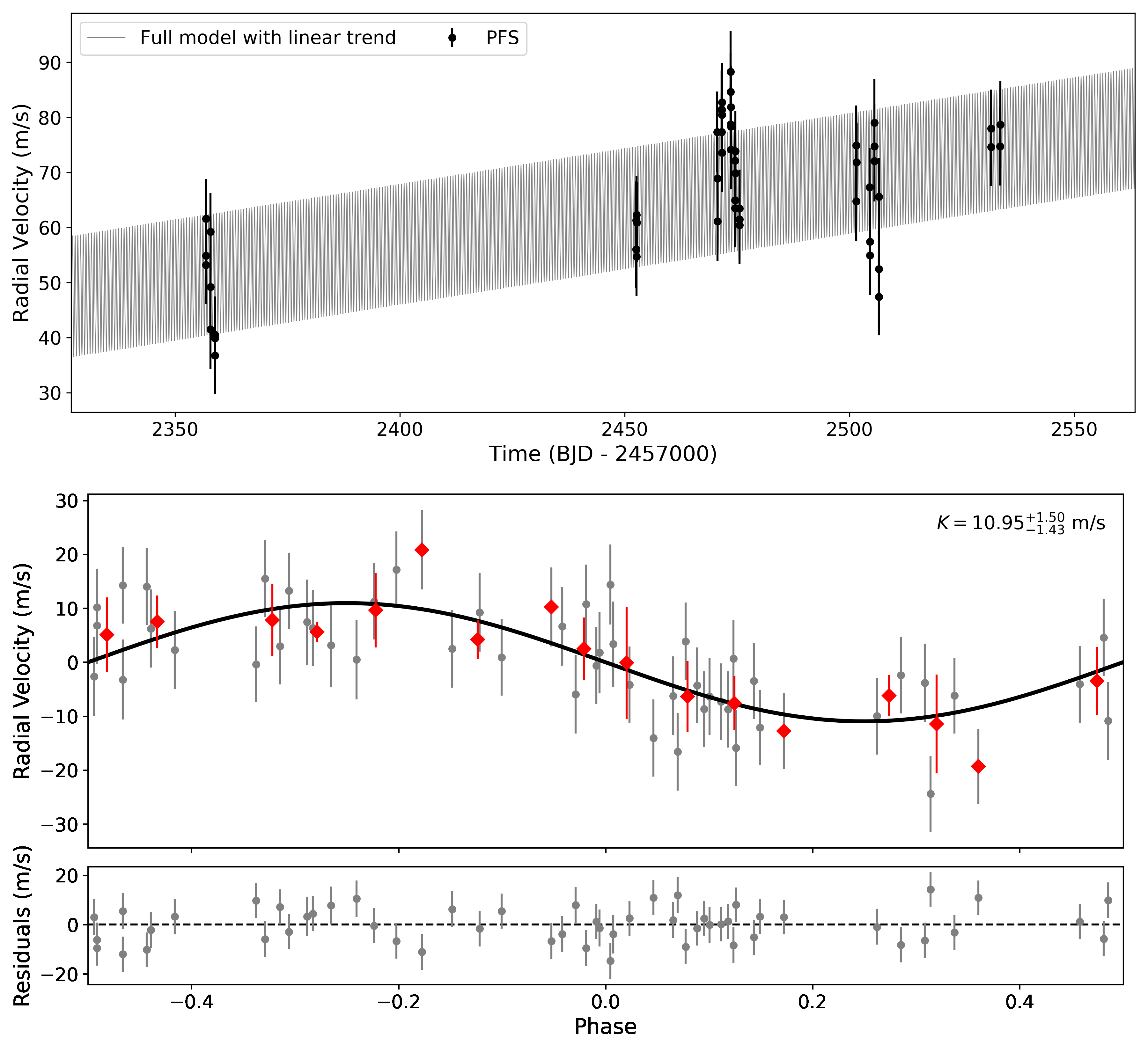}
\caption{Results of the \texttt{juliet} joint fit to the PFS radial velocities. \textit{Top:} PFS RVs over time (black points), and full best-fit \texttt{juliet} model (gray curve). \textit{Middle:} Phase-folded RV measurements from PFS. The black curve is the best-fit \texttt{juliet} RV model, the gray points are the unbinned RVs, and the red diamonds are the RV measurements binned in 0.05 units of orbital phase. Errorbars are the quadrature sum of the PFS internal uncertainties, and the RV jitter estimate from the \texttt{juliet} fit.The best-fit RV semi-amplitude is $K = 10.95^{+1.50}_{-1.43}~\ms$. \textit{Bottom:} RV residuals after the data have been subtracted from the best-fit model. \label{fig:RVsjointfit}}
\end{figure*}

We show the final transit and RV models of the joint fit based on the posterior sampling in Figures \ref{fig:transitsjointfit} and \ref{fig:RVsjointfit} respectively, the posterior parameters of our joint fit in Table \ref{tab:julietposteriors}, the selected priors for our joint fit in Table \ref{tab:julietpriors}, the obtained posterior probabilities in Figure \ref{fig:cornerposteriors}, and the derived planetary parameters of \thisplanetb\ based on the posteriors of the joint fit in Table \ref{tab:julietderived}.\\

To summarize, the \thisstar\ system consists of a late K-/early M-dwarf host star with at least one hot super-Earth planet, \thisplanetb\ (see Table \ref{tab:julietderived}), which has a mass of  $M_{p}$ = \plmassunc\ and radius of $R_{p}$ = \plradunc\, on a circular orbit with a period of 0.605~days. We derived a bulk density of $\rho$ = \plrhounc\ and an equilibrium temperature, assuming a zero albedo, of $T_{eq} = 1323\pm44$~K. 

\section{Discussion}
\label{sec:discussion}

Our RV measurements of \thisplanetb\ constrain the planetary mass with an uncertainty of $\sim14$\%, and the \TESS, LCOGT and MEarth-South light curves constrain the planetary radius with an uncertainty of $\sim7$\%. Thus, \thisplanetb\ belongs to the small group of super-Earths with precisely measured masses\footnote{Only planet masses measured by the radial velocity method are considered, in order to avoid differences in the planet mass distribution with other methods.} and radii, or those planets with measurement precision better than 30\% (see Figure \ref{fig:MR}). With precise mass and radius measurements in hand for \thisplanetb, we discuss implications for radius valley studies, the potential for planetary atmospheric characterization, potential planetary compositions, planetary bulk density, and possible planet formation mechanisms. We also discuss constraints on a potential second planet in the \thisstar\ system.


\subsection{Implications for the Radius Valley around M-dwarfs}

Several physical mechanisms/theoretical models have been proposed to explain the existence of the radius gap. These mechanisms include  thermally driven atmospheric mass loss, e.g., photoevaporation \citep{lopez2012thermal, chen2016evolutionary, owen2017evaporation} and core-powered mass loss \citep{ginzburg2018core,gupta2019sculpting,gupta2020signatures}; and gas-poor formation -- a natural outcome of planet formation, where rocky super-Earths are a result of formation in gas-poor environments, without requiring any atmospheric escape \citep{lee2014make,lee2016breeding,lopez2018formation,lee2021primordial}. The slope of the radius valley (in period-radius space) can be used to discern between a thermally driven mass-loss model or a gas-poor formation model \citep{lopez2018formation,gupta2020signatures,lee2021primordial}. 

The slope of the radius valley around Sun-like stars has been characterized using data from the \textit{Kepler} and \textit{K2} missions, and both thermally driven mass-loss and gas-poor formation models are favored in this stellar-mass regime \citep{van2018asteroseismic,martinez2019spectroscopic}. However, around low-mass mid-K- to mid-M-dwarfs, \citet{cloutier2020evolution} found tentative evidence that the slope of the radius valley is consistent with predictions from gas-poor formation. Additionally, the radius gap decreases as stellar radius decreases, and the radius gap is centered at $1.54\pm0.16$~R$_{\oplus }$ for low-mass K- and M-dwarf stars (vs.~1.75~R$_{\oplus }$ for Sun-like stars in \citealt{fulton2017california}). 

\thisplanetb\ lies between the predicted slopes of the thermally driven mass loss model and gas-poor formation model, and hence within the M-dwarf radius valley created by these mechanisms (the M-dwarf radius valley ranges from 1.5-2.5~R$_{\oplus }$ between the predicted slopes -- see Figure 9 in \citealp{cloutier2021toi1634}). \thisplanetb's orbital period (0.605~days) and size (\plradunc) make it a ``keystone planet" -- a valuable target to conduct tests of competing radius valley models across a range of stellar masses, using precise planetary mass and radius measurements \citep{cloutier2020evolution}. \thisplanetb\ joins TOI-1235~b \citep{cloutier2020toi1235, bluhm2020precise}, TOI-776~b \citep{luque2021planetary}, TOI-1685~b \citep{bluhm2021ultra}, and TOI-1634~b \citep{cloutier2021toi1634} as keystone planets that will help elucidate the physical mechanism that formed the radius valley around early M-dwarfs. Distinguishing between the two atmospheric loss mechanisms will require the discovery of additional keystone planets for statistical studies and population analysis, as well as atmospheric studies of \thisplanetb\ and other keystone planets to provide observational evidence to validate model predictions.


\subsection{Atmospheric Characterization Prospects \label{subsec:atmchr}}

Super-Earths with $R_{p} >$ 1.6~R$_{\earth}$ are expected to have a substantial H/He atmosphere \citep{rogers2015most}, and though \thisplanetb's radius (\plradunc) places it just above the radius gap, its bulk density is inconsistent with the presence of a low mean-molecular weight envelope. Based on \thisplanetb's predicted composition (see Section \ref{subsec:plcompdens}) and ultra-short orbital period, we do not expect the planet to have retained a H/He envelope. But, \thisplanetb\ could either have: no atmosphere (bare rock); a metal/silicate vapor atmosphere with a composition set by the vaporizing magma-ocean on the surface \citep{schaefer2009chemistry,ito2015theoretical} since \thisplanetb's equilibrium temperature is hot enough to melt a rocky surface \citep{mansfield2019identifying}; or, especially at the low-end of its allowed mean density range, possibly a thin H/He or CO$_2$ or other atmosphere. A detailed atmospheric model is needed to determine possible atmospheric compositions for \thisplanetb, which is beyond the scope of this work.

We calculated the emission spectroscopy metric (ESM) and transmission spectroscopy metric (TSM), as defined by \citet{kempton2018framework}, to determine \thisplanetb's potential for atmospheric characterization. Using the stellar parameters reported in Section \ref{sec:SED}, and the planetary parameters in Table \ref{tab:julietderived}, we obtain ESM~$= 10.1 \pm 1.6$ and TSM~$= 29 \pm 8$.

\thisplanetb\ is a good candidate for emission spectroscopy with the James Webb Space Telescope (\jwst). The planet may have a mineral-rich atmosphere consisting of metal and silicate vapors, since its equilibrium temperature is high enough to melt silicate rock \citep{schaefer2009chemistry, leger2011extreme, ito2015theoretical, ito2021hydrodynamic}. If \thisplanetb\ has no atmosphere, its surface may be characterized via secondary eclipse observations. With an ESM value of $10.1 \pm 1.6$, \thisplanetb\ is well above the threshold ESM of 7.5 suggested by \citet{kempton2018framework} for a high-quality atmospheric characterization target. \thisplanetb\ is one of only 8 super-Earths: 55~Cancri~e \citep{bourrier201855cnc}, HD~213885~b \citep{espinoza2020hd}, HD~3167~b \citep{christiansen2017three}, TOI-431~b \citep{osborn2021toi}, TOI-500~b \citep{serrano2022low}, TOI-1807~b \citep{nardiello2022gapstoi1807} and K2-141~b \citep{malavolta2018ultra}, with mass measurement precision $>5\sigma$ and with V/J/H/K $<$ 13 mag, that has an ESM $>$ 7.5. It is also the only super-Earth above the radius gap in the temperature range 1250--1750 K\footnote{https://tess.mit.edu/science/tess-acwg/ (as of 10/05/2022)}, which will allow us to probe an intermediate temperature range of hot super-Earths and explore the atmospheric species that have volatilized in this regime. 

In addition to having an ESM value above the suggested threshold, \thisplanetb\ can also be observed by \jwst\ for $\sim200$~days per year. Thus, \thisplanetb\ is accessible to \jwst\ for a significant portion of the year, which allows for more flexibility when planning observations.

\citet{kempton2018framework} suggests that TSM~$> 90$ be used as the threshold for planets with 1.5~R$_{\oplus }$ $<$ R$_{p}$ $<$ 10~R$_{\oplus }$. \thisplanetb\ does not meet this criteria because, due to its high mass and hence high surface gravity, it is unlikely to have an extended atmosphere that can be probed in transmission.


\subsection{Planetary Composition and Density \label{subsec:plcompdens}}

\begin{figure*}[htb!]
    \centering
    \includegraphics[width=\textwidth, keepaspectratio]{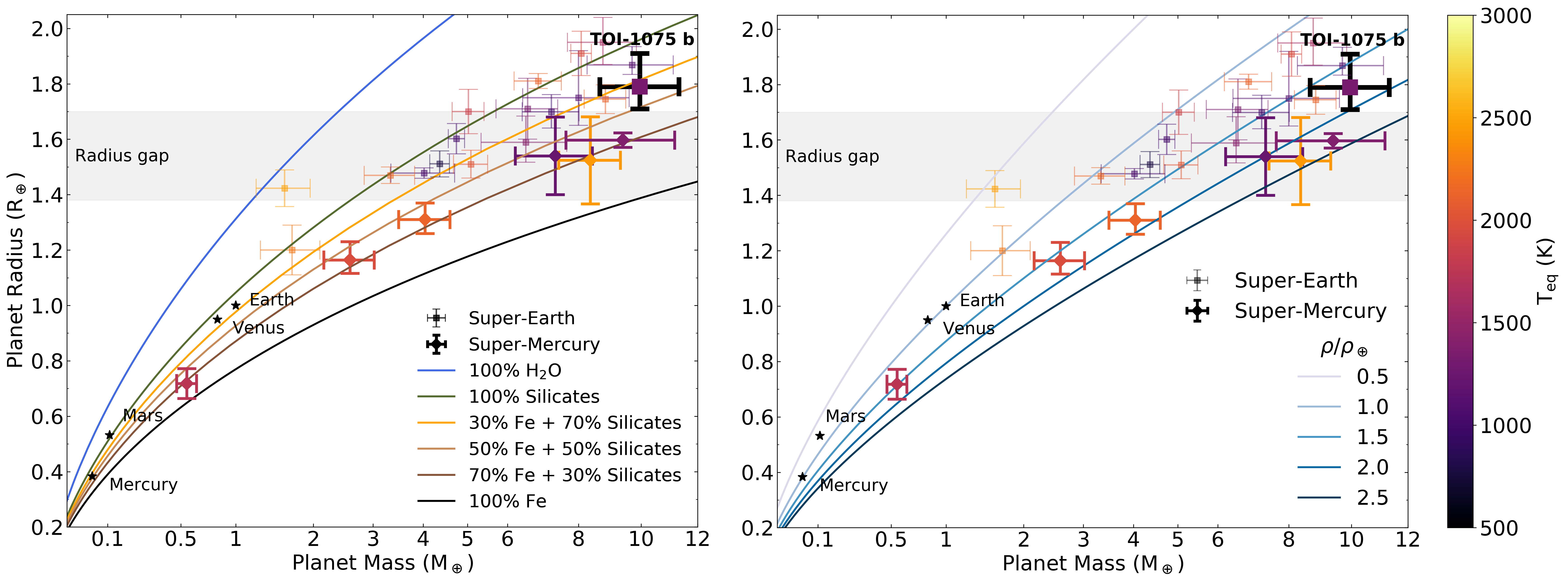}
\caption{Mass–radius diagram for small planets ($R_p$ $<$ 2~R$_{\earth}$ and $M_p$ $<$ 10~M$_{\earth}$) with measured mass and radius uncertainties below 30\%, as listed in the NASA Exoplanet Archive (only planet masses measured by the radial velocity method are considered). Planets are colored according to their calculated equilibrium temperature (assuming zero-albedo and efficient heat redistribution). Super-Earth planets have square symbols, and super-Mercury planets have bold diamond symbols. The gray shaded region denotes the radius gap for low-mass stars centered at $1.54\pm0.16$~R$_{\oplus }$ \citep{cloutier2020evolution}. The terrestrial solar system planets are plotted for reference. \textit{Left:} Mass-radius diagram with curves of constant composition. The solid lines are theoretical internal composition curves (\citealt{zeng_new_2021}; Lin et al. (in prep)) from top to bottom: 100\% H$_2$O (blue), 100\% silicates (green), a mixture of 30\% Fe and 70\%  silicates by mass (orange),  a mixture of 50\% Fe and 50\%  silicates by mass (light brown), a mixture of 70\% Fe and 30\%  silicates by mass (dark brown), and 100\% Fe (black). \thisplanetb's bulk density is consistent with a 35\% Fe + 65\% silicates by mass composition. \textit{Right:} Mass-radius diagram with curves of constant density relative to Earth. \thisplanetb's density is $\sim$1.75 times greater than the Earth. \label{fig:MR}}
\end{figure*}

The measured mass and radius of \thisplanetb\ result in a planetary bulk density of \plrhounc, which is almost twice as dense as the Earth ($\rho_{\earth}$ = 5.51~g cm$^{-3}$). Comparing \thisplanetb\ with the theoretical composition models of \citet{zeng_new_2021} and Lin et al. (in prep), the planet's bulk density is consistent with a 35\% Fe + 65\% silicates by mass composition (see Figure \ref{fig:MR}). 

To simulate \thisplanetb's interior, we numerically integrate three equations -- the mass of a spherical shell, hydrostatic equilibrium, and the equation of state (EOS) -- from the planet's center to the surface with a step size of 100~m using a planetary interior simulation code (Lin et al. (in prep)). The code is validated against recent mass-radius curves calculated by \cite{zeng_new_2021}. We assume a completely differentiated planet with an iron core and a mantle consisting of silicates. For iron and silicates, we adopt a second-order adapted polynomial EOS developed by \cite{holzapfel_coherent_2018}, using EOS coefficients listed in \cite{zeng_new_2021}. The inner boundary conditions of the simulated planets are assumed to be $M(0)=0$ and $P(0)=P_c$, where $P_c$ is the central pressure. We switch from the iron core to the silicate mantle when the desired core mass has been reached. The iteration terminates when the outer boundary condition $P(R)\le1$ bar is satisfied. We use a bisection method to search for the $P_c$ that produces $M(R)=M_p$ for a given core-mass fraction (CMF). Using the mean measured mass and radius of \thisplanetb, we compute a mean CMF of $0.35$. We then calculate a core-radius fraction (CRF) by dividing the radius of the iron layer by the total radius of the planet, resulting in a mean CRF of $0.52$, or $0.93\,$R$_{\earth}$. 

We further consider the most and least dense scenarios permissible by the mass and radius error bars. In the most dense scenario ($M_p = 11.31\,$M$_{\earth}$, $R_p = 1.709\,$R$_{\earth}$), we compute a CMF of 0.61 and a CRF of 0.67, or $1.15\,$R$_{\earth}$. In the least dense scenario ($M_p = 8.65\,$M$_{\earth}$, $R_p = 1.91\,$R$_{\earth}$), even a coreless pure silicates composition (CMF = 0, CRF = 0) cannot explain the low-end density of the planet. Even though such a coreless planet is unphysical from a planet formation perspective, we include this extreme scenario for completeness.

Though we have precisely measured the mass of \thisplanetb\ to $>7\sigma$ (\plmassunc), the uncertainty on the mass measurement leads to a wide range of possible CMF ($0.35^{+0.26}_{-0.35}$) and CRF ($0.52^{+0.15}_{-0.52}$) values, which can only be resolved with more precise mass and radius measurements not currently available. In the following sections, we discuss possible formation scenarios that could result in the mean density, most dense, and least dense scenarios. 

\subsubsection{\thisplanetb's Mean Density and Composition}

The mean mass and radius ($M_p = 9.95\,$M$_{\earth}$, $R_p = 1.791\,$R$_{\earth}$) of \thisplanetb\ result in a bulk density of 9.32~g~cm$^{-3}$. While \thisplanetb's mean bulk density is $\sim1.75\times$ that of Earth, its mean CMF (0.35) and CRF (0.52) are consistent with a predominantly rocky, Earth-like composition and internal structure. \thisplanetb's uncompressed density is 4.91~g~cm$^{-3}$, which is similar to the uncompressed density of the Earth (4.79~g~cm$^{-3}$), further supporting our findings of an Earth-like composition for \thisplanetb.

\thisplanetb\ is the most massive and densest super-Earth with 1.6~R$_{\oplus }$ $<$ R$_{p}$ $<$ 2~R$_{\oplus }$ discovered to date. The similarity in \thisplanetb's CMF relative to Earth (CMF$_{\earth}$ = 32.5\%, \citealp{seager2007mass}) supports our finding that \thisplanetb\ likely lacks a massive low mean-molecular weight envelope which, if present, would have corresponded to a larger observed radius for a given mass.

A handful of massive terrestrial planets like \thisplanetb\ have been discovered, but these objects are relatively rare. Planets similar in radius and Earth-like composition to \thisplanetb\ are listed in Table \ref{tab:massiveplanets}.

\subsubsection{\thisplanetb\ As A Super-Mercury}

In the most dense scenario ($M_p = 11.31\,$M$_{\earth}$, $R_p = 1.710\,$R$_{\earth}$, $\rho = 11.37$~g~cm$^{-3}$), we consider the possibility of \thisplanetb\ as a super-Mercury. The term ``Super-Mercuries" generally refers to planets that are super-Earth sized with enhanced uncompressed bulk densities, as compared to Earth-like planets. The high density has been interpreted to be indicative of a high iron-mass fraction, analogous to the solar system planet Mercury \citep{marcus2010minimum,adibekyan2021compositional}, and the formation mechanism for these planets is still unknown. Based on \thisplanetb's bulk density (\plrhounc), if the planet's mass is at the high range, with a high core-mass fraction in the allowed range (CMF = $0.35^{+0.26}_{-0.35}$), it could be a super-Mercury. The small group of recently identified super-Mercuries include: K2-38~b, K2-106~b, K2-229~b, Kepler-107~c, and Kepler-406~b \citep{adibekyan2021compositional}, as well as GJ~367~b \citep{lam2021gj} and HD~137496~b \citep{silva2022hd}, which are represented by bold data points in Figure \ref{fig:MR}. 

We investigated possible formation mechanisms that could result in the high density, and CMF and CRF ranges observed for \thisplanetb\ including: giant impacts \citep{marcus2010minimum, benz2008origin, asphaug2014mercury, liu2015giant, scora2020chemical, cambioni2021effect, cambioni2022metal}; in-situ formation \citep{weidenschilling1978iron, kruss2018seeding, johansen2022nucleation}; an initially metal-rich protoplanetary disk composition \citep{veyette2017physically, adibekyan2021compositional, schulze2021probability, souto2022detailed}; and the decompressed core of an evaporated gas giant \citep{hebrard2003evaporation, mocquet2014very}, but we do not have sufficient information to resolve the planet's formation mechanism.

\subsubsection{\thisplanetb\ As A Low-density Planet}

In the least dense scenario ($M_p = 8.65\,$M$_{\earth}$, $R_p = 1.907\,$R$_{\earth}$, $\rho = 7.47$~g~cm$^{-3}$), our interior model results in a coreless silicate planet, and requires an additional water layer to account for the inflated radius. Such a coreless planet is unphysical from a planet formation perspective -- a planet with a mass of \thisplanetb\ is expected to have fully differentiated e.g. \citet{rubie2015accretion,cambioni2021effect}. A water layer would be physically unstable at \thisplanetb's equilibrium temperature. Therefore, layers less dense than silicates must be added to the model to fit the minimum mass and maximum radius scenario. Possible candidates for these layers include low-pressure silicate phases, or a metal/silicate vapor atmosphere.

\begin{table*}[htb!]
    \caption{Massive terrestrial planet parameters}
    \centering
    \begin{tabular}{lcccl}
    \hline
    Planet Name & Planet Radius ($R_{\earth}$) & Planet Mass ($M_{\earth}$) & T$_{eq}$ (K) & Reference \\
     \hline
    \thisplanetb\ & $1.791_{-0.081}^{+0.116}$ & $9.95_{-1.30}^{+1.36}$ & 1323$\pm44$& This work\\
     Kepler-20~b & ${1.868}_{-0.034}^{+0.066}$ & ${9.70}_{-1.44}^{+1.41}$ & $1105\pm37$ & \citet{buchhave20161}\\
    LHS~1140~b & $1.727\pm0.032$ & $6.98\pm0.89$ & $235\pm5$ & \citet{ment2019second}\\
    TOI-1235~b & $1.738_{-0.076}^{+0.087}$ & ${6.91}_{-0.85}^{+0.75}$ & $754\pm18$ & \citet{cloutier2020toi1235}\\
    HD~213885~b & $1.74\pm0.05$ & $8.8\pm0.6$ & $2128\pm14$ & \citet{espinoza2020hd}\\
    WASP-47~e & $1.808\pm0.026$ & $6.77\pm0.57$ & $2514\pm70$ & \citet{bryant2022revisiting}\\
    K2-216~b & $1.72\pm0.06$ & $8.18\pm1.65$ & $1217\pm34$ & \citet{clark2022galah}\\
    \hline
    \label{tab:massiveplanets} \end{tabular}
\end{table*}


\subsection{A Potential Second Planet in the System \label{subsec:secondpl}}

We find a linear long-term trend in the PFS RV data (see Section 4.2) whose period is longer than the baseline of our RV observations. This may indicate the presence of a second planet in the system. We place lower limits on the period, semi-amplitude, and mass of a second planet candidate as the source of the long-term trend. 

The orbital period of the second planet candidate is at least twice the observing baseline of the PFS RV observations (otherwise we would have seen the RV trend turn over before our observations concluded as the potential planet passed through its quadrature phase). The PFS observations were taken over a period of $176.68$~days (T$_{\rm baseline}$), hence the orbital period of the planet candidate should be at least $353$~days. 

Taking the best-fit linear trend results from \texttt{juliet}, the RV data is shifted $0.130^{+0.017}_{-0.018}~\rm m\,s^{-1}\,day^{-1}$. Thus, the RV semi-amplitude of the planet candidate must be at least (T$_{\rm baseline}$ $\times$ RV Trend)/2 = $11.48~\ms$.

Combining the minimum orbital period ($353$~days) and the minimum RV semi-amplitude ($11.48~\ms$), and assuming a circular Keplarian orbit, the minimum mass of the second planet candidate is $\sim87~M_{\earth}$ or $\sim0.28~M_{\rm Jupiter}$. 

The presence of a second planet candidate motivates us to collect additional RV data for this system in order to determine the period and measure the mass of the second planet candidate, while also improving the uncertainty on the mass measurement of \thisplanetb. Additionally, there are currently a handful of systems consisting of a USP planet and a close-in companion, with periods ranging from a few days to tens of days e.g. K2-106 \citep{guenther2017k2}, K2-141 \citep{malavolta2018ultra}, K2-229 \citep{santerne2018earth}, TOI-500 \citep{serrano2022low}. These close-in companions could be responsible for migrating the USP planets to their current positions \citep{pu2019low, millholland2020formation, serrano2022low}. The possible existence of such a close-in companion in the \thisstar\ system serves as additional motivation for further RV follow-up of the system.

Determining the source of the long-term trend, and more accurately measuring \thisplanetb's planet mass and parameters will further detailed planet formation, planet migration and atmospheric characterization efforts, since a planet’s gravity plays an important role in its collisional history and interpreting atmospheric spectra \citep{batalha2019precision}.


\section{Conclusions\label{sec:conclusions}}

We report the discovery and confirmation of \thisplanetb, a transiting, ultra-short period, hot super-Earth orbiting a nearby ($d$ = 61.4 pc) late K-/early M-dwarf star. Using photometric observations from \TESS, LCOGT, and MEarth-South, and radial velocity observations from PFS, we precisely measure the radius and mass of \thisplanetb\ to be \plradunc\ and \plmassunc, respectively. Our PFS radial velocity data also suggest the presence of a second planet candidate in the system, with a minimum mass of $\sim87~M_{\earth}$ and a minimum orbital period of $\sim353$~days. \thisplanetb\ has a bulk density of \plrhounc, consistent with a composition of 35\% iron by mass, and a core-radius fraction of 52\%. \thisplanetb\ is a good candidate for emission spectroscopy with \jwst, which we can use to characterize a potentially mineral-rich atmosphere. \TESS\ is scheduled to observe \thisstar\ again in Year~5, Sector~67 (July 2023), which will provide a more precise planet radius, and the ability to search for variations in the planet period on a 4-year timescale. \thisplanetb\ is a massive, dense, high temperature, ultra-short period super-Earth inside of the M-dwarf radius valley, making the system ideal for testing planet formation and evolution theories, density enhancing mechanisms, and theoretical models related to atmospheric loss.

\clearpage
\startlongtable
\begin{deluxetable*}{llccc}
\tabletypesize{\small}
\tablewidth{\linewidth}
\tablecaption{Median values and 68\% confidence interval for posterior parameters from joint photometric and radial velocity \texttt{juliet} analysis for the TOI-1075 system.}
\tablehead{\colhead{Parameter} & \colhead{Units} & \colhead{Values}}
\startdata
\smallskip\\\multicolumn{2}{l}{Stellar Parameters:}&\smallskip\\
$\rho_*$ & Density (g cm$^{-3}$) & $3.73^{+0.99}_{-1.67}$\\
\smallskip\\\multicolumn{2}{l}{Planet Parameters:}&TOI-1075 b\smallskip\\
$P$ &Period (days) &$0.6047328\pm0.0000032$\\
$T_0$ & Time of transit center (\bjdtdb) &$2458654.2510^{+ 0.00040}_{-0.00050}$\\
$r_1$ & Parametrization of \citet{espinoza2018efficient} for $b$ & $0.60^{+0.20}_{-0.17}$\\
$r_2$ & Parametrization of \citet{espinoza2018efficient} for $R_p/R_*$ & $0.0282^{+0.0019}_{-0.0013}$\\
$K$ & RV semi-amplitude (\ms) & $10.95^{+1.50}_{-1.43}$\\
$e$ & Eccentricity (fixed) & $0.00$\\
\smallskip\\\multicolumn{2}{l}{Photometry Parameters:}\\
\multicolumn{2}{l}{}&MEarth-South&LCOGT&TESS\smallskip\\
$M$ & Relative flux offset & $0.0000$ & $0.0000$ & $-0.0001$\\
$\sigma$ &  Jitter term for light curve (ppm) & $901.29^{+86.80}_{-91.69}$ & $471.24^{+58.31}_{-62.17}$ & $3.21^{+25.84}_{-2.86}$\\
$q_{1}$ & Quadratic limb-darkening parametrization \citep{kipping2013efficient} & $0.54^{+0.28}_{-0.31}$ & $0.34^{+0.33}_{-0.23}$ & $0.59^{+0.24}_{-0.25}$\\
$q_{2}$ & Quadratic limb-darkening parametrization \citep{kipping2013efficient} & $0.47^{+0.33}_{-0.30}$ & $0.39^{+0.35}_{-0.27}$ & $0.47^{+0.29}_{-0.28}$\\
\smallskip\\\multicolumn{2}{l}{RV Parameters:}\smallskip\\
$\mu_{PFS}$ & Systemic velocity for PFS (\ms) & $-67.3^{+24.7}_{-20.1}$ \\
$\sigma_{PFS}$ &  Jitter term for PFS (\ms) & $6.60^{+0.81}_{-0.69}$ \\
$A$ & Slope of linear long-term RV trend (\ms~day$^{-1}$) & $0.130^{+0.017}_{-0.018}$ \\
$B$ & Intercept of linear long-term RV trend (\ms) & $-67.3^{+23.1}_{-20.6}$ \\
\label{tab:julietposteriors} \enddata
\end{deluxetable*}

\startlongtable
\begin{deluxetable*}{llccc}
\tabletypesize{\small}
\tablewidth{\linewidth}
\tablecaption{Median values and 68\% confidence interval for derived parameters from joint photometric and radial velocity \texttt{juliet} analysis for TOI-1075 b.}
\tablehead{\colhead{Parameter} & \colhead{Units} & \colhead{Values}}
\startdata
\smallskip\\\multicolumn{2}{l}{Derived Transit Parameters:}&\smallskip\\
$p=R_p/R_*$ & Radius of planet in stellar radii & $0.0282^{+0.0019}_{-0.0013}$\\
$a/R_*$ & Semi-major axis in stellar radii  & $4.40^{+0.89}_{-0.91}$\\
$b=(a/R_*)cos(i_p)$ & Transit impact parameter & $0.40^{+0.29}_{-0.25}$\\
$i_p$ & Inclination (degrees) & $84.67^{+3.34}_{-3.93}$\\
\smallskip\\\multicolumn{2}{l}{}&MEarth-South&LCOGT&TESS\smallskip\\
$u_{1}$ & Linear limb-darkening coefficient & $0.61^{+0.52}_{-0.40}$ & $0.39^{+0.46}_{-0.27}$ & $0.68\pm0.39$\\
$u_{2}$ & Quadratic limb-darkening coefficient & $0.035^{+0.043}_{-0.041}$ & $0.10^{+0.32}_{-0.37}$ & $0.042^{+0.044}_{-0.039}$\\
\smallskip\\\multicolumn{2}{l}{Derived Physical Parameters:}&TOI-1075 b\smallskip\\
$R_p$ & Radius (R$_{\earth}$) &$1.791^{+0.116}_{-0.081}$\\
$M_p$ & Mass (M$_{\earth}$) &$9.95^{+1.36}_{-1.30}$\\
$\rho_p$ & Density (g cm$^{-3}$) &$9.32^{+2.05}_{-1.85}$\\
$a$ & Semi-major axis (AU) &$0.01159^{+0.00023}_{-0.00020}$\\
$T_{eq}$ & Equilibrium temperature (K) & $1323\pm44$\\
$g_p$ & Surface gravity (m~s$^{-2}$) & $30.0^{+5.4}_{-4.8}$\\
$S_{p}$ &Insolation (S$_{\earth}$) & $509.2^{+18.7}_{-20.4}$\\
\label{tab:julietderived} \enddata
\end{deluxetable*}


\acknowledgements

This paper includes data collected by the \TESS\ mission. Funding for the \TESS\ mission is provided by NASA's Science Mission directorate. 
%
This paper includes data gathered with the 6.5 meter Magellan Telescopes located at Las Campanas Observatory, Chile. 
This work makes use of observations from the LCOGT network. Part of the LCOGT telescope time was granted by NOIRLab through the Mid-Scale Innovations Program (MSIP). MSIP is funded by NSF.
This work has made use of data from the European Space Agency (ESA) mission Gaia (https://www.cosmos.esa.int/gaia), processed by the Gaia Data Processing and Analysis Consortium (DPAC, https://www.cosmos.esa.int/web/gaia/dpac/consortium). Funding for the DPAC has been provided by national institutions, in particular the institutions participating in the Gaia Multilateral Agreement. 
Some of the observations in the paper made use of the High-Resolution Imaging instrument Zorro obtained under Gemini LLP Proposal Number: GN/S-2021A-LP-105. Zorro was funded by the NASA Exoplanet Exploration Program and built at the NASA Ames Research Center by Steve B. Howell, Nic Scott, Elliott P. Horch, and Emmett Quigley. Zorro was mounted on the Gemini North (and/or South) telescope of the international Gemini Observatory, a program of NSF’s OIR Lab, which is managed by the Association of Universities for Research in Astronomy (AURA) under a cooperative agreement with the National Science Foundation. on behalf of the Gemini partnership: the National Science Foundation (United States), National Research Council (Canada), Agencia Nacional de Investigación y Desarrollo (Chile), Ministerio de Ciencia, Tecnología e Innovación (Argentina), Ministério da Ciência, Tecnologia, Inovações e Comunicações (Brazil), and Korea Astronomy and Space Science Institute (Republic of Korea).
The MEarth Team gratefully acknowledges funding from the David and Lucile Packard Fellowship for Science and Engineering (awarded to D.C.). This material is based upon work supported by the National Science Foundation under grants AST-0807690, AST-1109468, AST-1004488 (Alan T. Waterman Award), and AST-1616624, and upon work supported by the National Aeronautics and Space Administration under Grant No. 80NSSC18K0476 issued through the XRP Program. This work is made possible by a grant from the John Templeton Foundation. The opinions expressed in this publication are those of the authors and do not necessarily reflect the views of the John Templeton Foundation.
Resources supporting this work were provided by the NASA High-End Computing (HEC) Program through the NASA Advanced Supercomputing (NAS) Division at Ames Research Center for the production of the SPOC data products. 
This research was carried out in part at the Jet Propulsion Laboratory, California Institute of Technology, under a contract with the National Aeronautics and Space Administration (80NM0018D0004).
Some of the data presented in this paper were obtained from the Mikulski Archive for Space Telescopes (MAST). The specific observations analyzed can be accessed via \dataset[https://doi.org/10.17909/t9-wpz1-8s54]{https://doi.org/10.17909/t9-wpz1-8s54}.
This research has made use of the NASA Exoplanet Archive \citep{planetarysystemsexoarchive}, made available by the NASA Exoplanet Science Institute at IPAC, which is operated by the California Institute of Technology under contract with the National Aeronautics and Space Administration.
This research has made use of NASA's Astrophysics Data System. 
This research has made use of the Exoplanet Follow-up Observation Program (ExoFOP; \citealp{https://doi.org/10.26134/exofop5}) website, which is operated by the California Institute of Technology, under contract with the National Aeronautics and Space Administration under the Exoplanet Exploration Program."
This research made use of Astropy, a community-developed core Python package for Astronomy \citep{Astropy2013}. 
This publication makes use of VOSA, developed under the Spanish Virtual Observatory (https://svo.cab.inta-csic.es) project funded by MCIN/AEI/10.13039/501100011033/ through grant PID2020-112949GB-I00.
VOSA has been partially updated by using funding from the European Union's Horizon 2020 Research and Innovation Programme, under Grant Agreement nº 776403 (EXOPLANETS-A). 
%


\facilities{TESS, Magellan:Clay (Planet Finder Spectrograph), Gemini-South (Zorro), SOAR, LCOGT, MEarth} 

\software{{AstroImageJ \citep{Collins:2017}, Astropy \citep{Astropy2013}, TAPIR \citep{Jensen:2013}, juliet \citep{espinoza2019juliet}, dynesty \citep{speagle2020dynesty}}}

\bibliographystyle{apj}
\bibliography{main}

\begin{thebibliography}{}
\expandafter\ifx\csname natexlab\endcsname\relax\def\natexlab#1{#1}\fi

\bibitem[{ESA(1997)}]{ESA1997}
 1997, ESA Special Publication, Vol. 1200, {The HIPPARCOS and TYCHO catalogues.
  Astrometric and photometric star catalogues derived from the ESA HIPPARCOS
  Space Astrometry Mission}

\bibitem[{Adams {et~al.}(2016)Adams, Jackson, \& Endl}]{adams2016ultra}
Adams, E.~R., Jackson, B., \& Endl, M. 2016, The Astronomical Journal, 152, 47

\bibitem[{Adams {et~al.}(2017)Adams, Jackson, Endl, Cochran, MacQueen, Duev,
  Jensen-Clem, Salama, Ziegler, Baranec, {et~al.}}]{adams2017ultra}
Adams, E.~R., Jackson, B., Endl, M., {et~al.} 2017, The Astronomical Journal,
  153, 82

\bibitem[{Adams {et~al.}(2021)Adams, Jackson, Johnson, Ciardi, Cochran, Endl,
  Everett, Furlan, Howell, Jayanthi, {et~al.}}]{adams2021ultra}
Adams, E.~R., Jackson, B., Johnson, S., {et~al.} 2021, The Planetary Science
  Journal, 2, 152

\bibitem[{Adibekyan {et~al.}(2021)Adibekyan, Dorn, Sousa, Santos, Bitsch,
  Israelian, Mordasini, Barros, Delgado~Mena, Demangeon,
  {et~al.}}]{adibekyan2021compositional}
Adibekyan, V., Dorn, C., Sousa, S.~G., {et~al.} 2021, Science, 374, 330

\bibitem[{Asphaug \& Reufer(2014)}]{asphaug2014mercury}
Asphaug, E., \& Reufer, A. 2014, Nature Geoscience, 7, 564

\bibitem[{{Astropy Collaboration} {et~al.}(2013){Astropy Collaboration},
  {Robitaille}, {Tollerud}, {Greenfield}, {Droettboom}, {Bray}, {Aldcroft},
  {Davis}, {Ginsburg}, {Price-Whelan}, {Kerzendorf}, {Conley}, {Crighton},
  {Barbary}, {Muna}, {Ferguson}, {Grollier}, {Parikh}, {Nair}, {Unther},
  {Deil}, {Woillez}, {Conseil}, {Kramer}, {Turner}, {Singer}, {Fox}, {Weaver},
  {Zabalza}, {Edwards}, {Azalee Bostroem}, {Burke}, {Casey}, {Crawford},
  {Dencheva}, {Ely}, {Jenness}, {Labrie}, {Lim}, {Pierfederici}, {Pontzen},
  {Ptak}, {Refsdal}, {Servillat}, \& {Streicher}}]{Astropy2013}
{Astropy Collaboration}, {Robitaille}, T.~P., {Tollerud}, E.~J., {et~al.} 2013,
  \aap, 558, A33

\bibitem[{{Bailer-Jones} {et~al.}(2021){Bailer-Jones}, {Rybizki}, {Fouesneau},
  {Demleitner}, \& {Andrae}}]{BailerJones2021}
{Bailer-Jones}, C.~A.~L., {Rybizki}, J., {Fouesneau}, M., {Demleitner}, M., \&
  {Andrae}, R. 2021, \aj, 161, 147

\bibitem[{Batalha {et~al.}(2019)Batalha, Lewis, Fortney, Batalha, Kempton,
  Lewis, \& Line}]{batalha2019precision}
Batalha, N.~E., Lewis, T., Fortney, J.~J., {et~al.} 2019, The Astrophysical
  Journal Letters, 885, L25

\bibitem[{Batalha(2014)}]{batalha2014exploring}
Batalha, N.~M. 2014, Proceedings of the National Academy of Sciences, 111,
  12647

\bibitem[{{Bensby} {et~al.}(2014){Bensby}, {Feltzing}, \& {Oey}}]{Bensby2014}
{Bensby}, T., {Feltzing}, S., \& {Oey}, M.~S. 2014, \aap, 562, A71

\bibitem[{Benz {et~al.}(2008)Benz, Anic, Horner, \& Whitby}]{benz2008origin}
Benz, W., Anic, A., Horner, J., \& Whitby, J.~A. 2008, in Mercury (Springer),
  7--20

\bibitem[{Berta {et~al.}(2012)Berta, Irwin, Charbonneau, Burke, \&
  Falco}]{berta2012transit}
Berta, Z.~K., Irwin, J., Charbonneau, D., Burke, C.~J., \& Falco, E.~E. 2012,
  The Astronomical Journal, 144, 145

\bibitem[{Bluhm {et~al.}(2020)Bluhm, Luque, Espinoza, Pall{\'e}, Caballero,
  Dreizler, Livingston, Mathur, Quirrenbach, Stock,
  {et~al.}}]{bluhm2020precise}
Bluhm, P., Luque, R., Espinoza, N., {et~al.} 2020, Astronomy \& Astrophysics,
  639, A132

\bibitem[{Bluhm {et~al.}(2021)Bluhm, Pall{\'e}, Molaverdikhani, Kemmer, Hatzes,
  Kossakowski, Stock, Caballero, Lillo-Box, B{\'e}jar,
  {et~al.}}]{bluhm2021ultra}
Bluhm, P., Pall{\'e}, E., Molaverdikhani, K., {et~al.} 2021, Astronomy \&
  Astrophysics, 650, A78

\bibitem[{{Boro Saikia} {et~al.}(2018){Boro Saikia}, {Marvin}, {Jeffers},
  {Reiners}, {Cameron}, {Marsden}, {Petit}, {Warnecke}, \&
  {Yadav}}]{BoroSaikia2018}
{Boro Saikia}, S., {Marvin}, C.~J., {Jeffers}, S.~V., {et~al.} 2018, \aap, 616,
  A108

\bibitem[{Borucki {et~al.}(2010)Borucki, Koch, Basri, Batalha, Brown, Caldwell,
  Caldwell, Christensen-Dalsgaard, Cochran, DeVore, Dunham, Dupree, Gautier,
  Geary, Gilliland, Gould, Howell, Jenkins, Kondo, Latham, Marcy, Meibom,
  Kjeldsen, Lissauer, Monet, Morrison, Sasselov, Tarter, Boss, Brownlee, Owen,
  Buzasi, Charbonneau, Doyle, Fortney, Ford, Holman, Seager, Steffen, Welsh,
  Rowe, Anderson, Buchhave, Ciardi, Walkowicz, Sherry, Horch, Isaacson,
  Everett, Fischer, Torres, Johnson, Endl, MacQueen, Bryson, Dotson, Haas,
  Kolodziejczak, Cleve, Chandrasekaran, Twicken, Quintana, Clarke, Allen, Li,
  Wu, Tenenbaum, Verner, Bruhweiler, Barnes, \& Prsa}]{borucki2010kepler}
Borucki, W.~J., Koch, D., Basri, G., {et~al.} 2010, Science, 327, 977

\bibitem[{Bourrier {et~al.}(2018)Bourrier, Dumusque, Dorn, Henry,
  Astudillo-Defru, Rey, Benneke, H{\'e}brard, Lovis, Demory,
  {et~al.}}]{bourrier201855cnc}
Bourrier, V., Dumusque, X., Dorn, C., {et~al.} 2018, Astronomy \& Astrophysics,
  619, A1

\bibitem[{{Boyajian} {et~al.}(2012){Boyajian}, {von Braun}, {van Belle},
  {McAlister}, {ten Brummelaar}, {Kane}, {Muirhead}, {Jones}, {White},
  {Schaefer}, {Ciardi}, {Henry}, {L{\'o}pez-Morales}, {Ridgway}, {Gies}, {Jao},
  {Rojas-Ayala}, {Parks}, {Sturmann}, {Sturmann}, {Turner}, {Farrington},
  {Goldfinger}, \& {Berger}}]{Boyajian:2012}
{Boyajian}, T.~S., {von Braun}, K., {van Belle}, G., {et~al.} 2012, \apj, 757,
  112

\bibitem[{{Brown} {et~al.}(2013){Brown}, {Baliber}, {Bianco}, {Bowman},
  {Burleson}, {Conway}, {Crellin}, {Depagne}, {De Vera}, {Dilday}, {Dragomir},
  {Dubberley}, {Eastman}, {Elphick}, {Falarski}, {Foale}, {Ford}, {Fulton},
  {Garza}, {Gomez}, {Graham}, {Greene}, {Haldeman}, {Hawkins}, {Haworth},
  {Haynes}, {Hidas}, {Hjelstrom}, {Howell}, {Hygelund}, {Lister}, {Lobdill},
  {Martinez}, {Mullins}, {Norbury}, {Parrent}, {Paulson}, {Petry}, {Pickles},
  {Posner}, {Rosing}, {Ross}, {Sand}, {Saunders}, {Shobbrook}, {Shporer},
  {Street}, {Thomas}, {Tsapras}, {Tufts}, {Valenti}, {Vander Horst}, {Walker},
  {White}, \& {Willis}}]{Brown:2013}
{Brown}, T.~M., {Baliber}, N., {Bianco}, F.~B., {et~al.} 2013, \pasp, 125, 1031

\bibitem[{Bryant \& Bayliss(2022)}]{bryant2022revisiting}
Bryant, E.~M., \& Bayliss, D. 2022, The Astronomical Journal, 163, 197

\bibitem[{Buchhave {et~al.}(2016)Buchhave, Dressing, Dumusque, Rice,
  Vanderburg, Mortier, Lopez-Morales, Lopez, Lundkvist, Kjeldsen,
  {et~al.}}]{buchhave20161}
Buchhave, L.~A., Dressing, C.~D., Dumusque, X., {et~al.} 2016, The Astronomical
  Journal, 152, 160

\bibitem[{{Buder} {et~al.}(2018){Buder}, {Asplund}, {Duong}, {Kos}, {Lind},
  {Ness}, {Sharma}, {Bland-Hawthorn}, {Casey}, {de Silva}, {D'Orazi},
  {Freeman}, {Lewis}, {Lin}, {Martell}, {Schlesinger}, {Simpson}, {Zucker},
  {Zwitter}, {Amarsi}, {Anguiano}, {Carollo}, {Casagrande}, {{\v{C}}otar},
  {Cottrell}, {da Costa}, {Gao}, {Hayden}, {Horner}, {Ireland}, {Kafle},
  {Munari}, {Nataf}, {Nordlander}, {Stello}, {Ting}, {Traven}, {Watson},
  {Wittenmyer}, {Wyse}, {Yong}, {Zinn}, {{\v{Z}}erjal}, \& {Galah
  Collaboration}}]{Buder2018}
{Buder}, S., {Asplund}, M., {Duong}, L., {et~al.} 2018, \mnras, 478, 4513

\bibitem[{{Buder} {et~al.}(2021){Buder}, {Sharma}, {Kos}, {Amarsi},
  {Nordlander}, {Lind}, {Martell}, {Asplund}, {Bland-Hawthorn}, {Casey}, {de
  Silva}, {D'Orazi}, {Freeman}, {Hayden}, {Lewis}, {Lin}, {Schlesinger},
  {Simpson}, {Stello}, {Zucker}, {Zwitter}, {Beeson}, {Buck}, {Casagrande},
  {Clark}, {{\v{C}}otar}, {da Costa}, {de Grijs}, {Feuillet}, {Horner},
  {Kafle}, {Khanna}, {Kobayashi}, {Liu}, {Montet}, {Nandakumar}, {Nataf},
  {Ness}, {Spina}, {Tepper-Garc{\'\i}a}, {Ting}, {Traven},
  {Vogrin{\v{c}}i{\v{c}}}, {Wittenmyer}, {Wyse}, {{\v{Z}}erjal}, \& {GALAH
  Collaboration}}]{Buder2021}
{Buder}, S., {Sharma}, S., {Kos}, J., {et~al.} 2021, \mnras, 506, 150

\bibitem[{{Butler} {et~al.}(1996){Butler}, {Marcy}, {Williams}, {McCarthy},
  {Dosanjh}, \& {Vogt}}]{butler1996}
{Butler}, R.~P., {Marcy}, G.~W., {Williams}, E., {et~al.} 1996, \pasp, 108, 500

\bibitem[{Cambioni {et~al.}(2022)Cambioni, Asphaug, Jung, Emsenhuber, \&
  Weiss}]{cambioni2022metal}
Cambioni, S., Asphaug, E., Jung, E., Emsenhuber, A., \& Weiss, B. 2022, LPI
  Contributions, 2678, 1979

\bibitem[{Cambioni {et~al.}(2021)Cambioni, Jacobson, Emsenhuber, Asphaug,
  Rubie, Gabriel, Schwartz, \& Furfaro}]{cambioni2021effect}
Cambioni, S., Jacobson, S.~A., Emsenhuber, A., {et~al.} 2021, The Planetary
  Science Journal, 2, 93

\bibitem[{{Casagrande} {et~al.}(2011){Casagrande}, {Sch{\"o}nrich}, {Asplund},
  {Cassisi}, {Ram{\'\i}rez}, {Mel{\'e}ndez}, {Bensby}, \&
  {Feltzing}}]{Casagrande2011}
{Casagrande}, L., {Sch{\"o}nrich}, R., {Asplund}, M., {et~al.} 2011, \aap, 530,
  A138

\bibitem[{Chen \& Rogers(2016)}]{chen2016evolutionary}
Chen, H., \& Rogers, L.~A. 2016, The Astrophysical Journal, 831, 180

\bibitem[{Christiansen {et~al.}(2017)Christiansen, Vanderburg, Burt, Fulton,
  Batygin, Benneke, Brewer, Charbonneau, Ciardi, Cameron,
  {et~al.}}]{christiansen2017three}
Christiansen, J.~L., Vanderburg, A., Burt, J., {et~al.} 2017, The Astronomical
  Journal, 154, 122

\bibitem[{Ciardi {et~al.}(2015)Ciardi, Beichman, Horch, \&
  Howell}]{ciardi2015understanding}
Ciardi, D.~R., Beichman, C.~A., Horch, E.~P., \& Howell, S.~B. 2015, The
  Astrophysical Journal, 805, 16

\bibitem[{Clark {et~al.}(2022)Clark, Wright, Wittenmyer, Horner, Hinkel,
  Clert{\'e}, Carter, Buder, Hayden, Bland-Hawthorn, {et~al.}}]{clark2022galah}
Clark, J.~T., Wright, D.~J., Wittenmyer, R.~A., {et~al.} 2022, Monthly Notices
  of the Royal Astronomical Society, 510, 2041

\bibitem[{Cloutier \& Menou(2020)}]{cloutier2020evolution}
Cloutier, R., \& Menou, K. 2020, The Astronomical Journal, 159, 211

\bibitem[{Cloutier {et~al.}(2020)Cloutier, Rodriguez, Irwin, Charbonneau,
  Stassun, Mortier, Latham, Isaacson, Howard, Udry,
  {et~al.}}]{cloutier2020toi1235}
Cloutier, R., Rodriguez, J.~E., Irwin, J., {et~al.} 2020, The Astronomical
  Journal, 160, 22

\bibitem[{Cloutier {et~al.}(2021)Cloutier, Charbonneau, Stassun, Murgas,
  Mortier, Massey, Lissauer, Latham, Irwin, Haywood,
  {et~al.}}]{cloutier2021toi1634}
Cloutier, R., Charbonneau, D., Stassun, K.~G., {et~al.} 2021, The Astronomical
  Journal, 162, 79

\bibitem[{{Collins} {et~al.}(2017){Collins}, {Kielkopf}, {Stassun}, \&
  {Hessman}}]{Collins:2017}
{Collins}, K.~A., {Kielkopf}, J.~F., {Stassun}, K.~G., \& {Hessman}, F.~V.
  2017, \aj, 153, 77

\bibitem[{{Crane} {et~al.}(2006){Crane}, {Shectman}, \& {Butler}}]{crane2006}
{Crane}, J.~D., {Shectman}, S.~A., \& {Butler}, R.~P. 2006, in Society of
  Photo-Optical Instrumentation Engineers (SPIE) Conference Series, Vol. 6269,
  Society of Photo-Optical Instrumentation Engineers (SPIE) Conference Series,
  ed. I.~S. {McLean} \& M.~{Iye}, 626931

\bibitem[{{Crane} {et~al.}(2010){Crane}, {Shectman}, {Butler}, {Thompson},
  {Birk}, {Jones}, \& {Burley}}]{crane2010}
{Crane}, J.~D., {Shectman}, S.~A., {Butler}, R.~P., {et~al.} 2010, in Society
  of Photo-Optical Instrumentation Engineers (SPIE) Conference Series, Vol.
  7735, Ground-based and Airborne Instrumentation for Astronomy III, ed. I.~S.
  {McLean}, S.~K. {Ramsay}, \& H.~{Takami}, 773553

\bibitem[{{Crane} {et~al.}(2008){Crane}, {Shectman}, {Butler}, {Thompson}, \&
  {Burley}}]{crane2008}
{Crane}, J.~D., {Shectman}, S.~A., {Butler}, R.~P., {Thompson}, I.~B., \&
  {Burley}, G.~S. 2008, in Society of Photo-Optical Instrumentation Engineers
  (SPIE) Conference Series, Vol. 7014, Ground-based and Airborne
  Instrumentation for Astronomy II, ed. I.~S. {McLean} \& M.~M. {Casali},
  701479

\bibitem[{{Cutri} \& {et al.}(2012)}]{Cutri2012}
{Cutri}, R.~M., \& {et al.} 2012, VizieR Online Data Catalog, II/311

\bibitem[{{Cutri} {et~al.}(2003){Cutri}, {Skrutskie}, {van Dyk}, {Beichman},
  {Carpenter}, {Chester}, {Cambresy}, {Evans}, {Fowler}, {Gizis}, {Howard},
  {Huchra}, {Jarrett}, {Kopan}, {Kirkpatrick}, {Light}, {Marsh}, {McCallon},
  {Schneider}, {Stiening}, {Sykes}, {Weinberg}, {Wheaton}, {Wheelock}, \&
  {Zacarias}}]{Cutri2003}
{Cutri}, R.~M., {Skrutskie}, M.~F., {van Dyk}, S., {et~al.} 2003, {2MASS All
  Sky Catalog of point sources.}

\bibitem[{Dressing {et~al.}(2015)Dressing, Charbonneau, Dumusque, Gettel, Pepe,
  Cameron, Latham, Molinari, Udry, Affer, Bonomo, Buchhave, Cosentino,
  Figueira, Fiorenzano, Harutyunyan, Haywood, Johnson, Lopez-Morales, Lovis,
  Malavolta, Mayor, Micela, Motalebi, Nascimbeni, Phillips, Piotto, Pollacco,
  Queloz, Rice, Sasselov, S{\'{e}}gransan, Sozzetti, Szentgyorgyi, \&
  Watson}]{dressing2015mass}
Dressing, C.~D., Charbonneau, D., Dumusque, X., {et~al.} 2015, The
  Astrophysical Journal, 800, 135

\bibitem[{{Egeland} {et~al.}(2017){Egeland}, {Soon}, {Baliunas}, {Hall},
  {Pevtsov}, \& {Bertello}}]{Egeland2017}
{Egeland}, R., {Soon}, W., {Baliunas}, S., {et~al.} 2017, \apj, 835, 25

\bibitem[{Espinoza(2018)}]{espinoza2018efficient}
Espinoza, N. 2018, arXiv preprint arXiv:1811.04859

\bibitem[{Espinoza {et~al.}(2019)Espinoza, Kossakowski, \&
  Brahm}]{espinoza2019juliet}
Espinoza, N., Kossakowski, D., \& Brahm, R. 2019, Monthly Notices of the Royal
  Astronomical Society, 490, 2262

\bibitem[{Espinoza {et~al.}(2020)Espinoza, Brahm, Henning, Jord{\'a}n, Dorn,
  Rojas, Sarkis, Kossakowski, Schlecker, D{\'\i}az, {et~al.}}]{espinoza2020hd}
Espinoza, N., Brahm, R., Henning, T., {et~al.} 2020, Monthly Notices of the
  Royal Astronomical Society, 491, 2982

\bibitem[{{Fantin} {et~al.}(2019){Fantin}, {C{\^o}t{\'e}}, {McConnachie},
  {Bergeron}, {Cuillandre}, {Gwyn}, {Ibata}, {Thomas}, {Carlberg}, {Fabbro},
  {Haywood}, {Lan{\c{c}}on}, {Lewis}, {Malhan}, {Martin}, {Navarro}, {Scott},
  \& {Starkenburg}}]{Fantin2019}
{Fantin}, N.~J., {C{\^o}t{\'e}}, P., {McConnachie}, A.~W., {et~al.} 2019, \apj,
  887, 148

\bibitem[{{Fuhrmann} {et~al.}(2017){Fuhrmann}, {Chini}, {Kaderhandt}, \&
  {Chen}}]{Fuhrmann2017}
{Fuhrmann}, K., {Chini}, R., {Kaderhandt}, L., \& {Chen}, Z. 2017, \mnras, 464,
  2610

\bibitem[{Fulton {et~al.}(2018)Fulton, Petigura, Blunt, \&
  Sinukoff}]{fulton2018radvel}
Fulton, B.~J., Petigura, E.~A., Blunt, S., \& Sinukoff, E. 2018, Publications
  of the Astronomical Society of the Pacific, 130, 044504

\bibitem[{Fulton {et~al.}(2017)Fulton, Petigura, Howard, Isaacson, Marcy,
  Cargile, Hebb, Weiss, Johnson, Morton, {et~al.}}]{fulton2017california}
Fulton, B.~J., Petigura, E.~A., Howard, A.~W., {et~al.} 2017, The Astronomical
  Journal, 154, 109

\bibitem[{{Gagn{\'e}} {et~al.}(2018){Gagn{\'e}}, {Mamajek}, {Malo}, {Riedel},
  {Rodriguez}, {Lafreni{\`e}re}, {Faherty}, {Roy-Loubier}, {Pueyo}, {Robin}, \&
  {Doyon}}]{Gagne2018}
{Gagn{\'e}}, J., {Mamajek}, E.~E., {Malo}, L., {et~al.} 2018, \apj, 856, 23

\bibitem[{{Gaia Collaboration} {et~al.}(2018){Gaia Collaboration}, {Brown},
  {Vallenari}, {Prusti}, {de Bruijne}, {Babusiaux}, {Bailer-Jones}, {Biermann},
  {Evans}, {Eyer}, {Jansen}, {Jordi}, {Klioner}, {Lammers}, {Lindegren},
  {Luri}, {Mignard}, {Panem}, {Pourbaix}, {Randich}, {Sartoretti}, {Siddiqui},
  {Soubiran}, {van Leeuwen}, {Walton}, {Arenou}, {Bastian}, {Cropper},
  {Drimmel}, {Katz}, {Lattanzi}, {Bakker}, {Cacciari}, {Casta{\~n}eda},
  {Chaoul}, {Cheek}, {De Angeli}, {Fabricius}, {Guerra}, {Holl}, {Masana},
  {Messineo}, {Mowlavi}, {Nienartowicz}, {Panuzzo}, {Portell}, {Riello},
  {Seabroke}, {Tanga}, {Th{\'e}venin}, {Gracia-Abril}, {Comoretto},
  {Garcia-Reinaldos}, {Teyssier}, {Altmann}, {Andrae}, {Audard},
  {Bellas-Velidis}, {Benson}, {Berthier}, {Blomme}, {Burgess}, {Busso},
  {Carry}, {Cellino}, {Clementini}, {Clotet}, {Creevey}, {Davidson}, {De
  Ridder}, {Delchambre}, {Dell'Oro}, {Ducourant},
  {Fern{\'a}ndez-Hern{\'a}ndez}, {Fouesneau}, {Fr{\'e}mat}, {Galluccio},
  {Garc{\'\i}a-Torres}, {Gonz{\'a}lez-N{\'u}{\~n}ez}, {Gonz{\'a}lez-Vidal},
  {Gosset}, {Guy}, {Halbwachs}, {Hambly}, {Harrison}, {Hern{\'a}ndez},
  {Hestroffer}, {Hodgkin}, {Hutton}, {Jasniewicz}, {Jean-Antoine-Piccolo},
  {Jordan}, {Korn}, {Krone-Martins}, {Lanzafame}, {Lebzelter}, {L{\"o}ffler},
  {Manteiga}, {Marrese}, {Mart{\'\i}n-Fleitas}, {Moitinho}, {Mora}, {Muinonen},
  {Osinde}, {Pancino}, {Pauwels}, {Petit}, {Recio-Blanco}, {Richards},
  {Rimoldini}, {Robin}, {Sarro}, {Siopis}, {Smith}, {Sozzetti}, {S{\"u}veges},
  {Torra}, {van Reeven}, {Abbas}, {Abreu Aramburu}, {Accart}, {Aerts},
  {Altavilla}, {{\'A}lvarez}, {Alvarez}, {Alves}, {Anderson}, {Andrei},
  {Anglada Varela}, {Antiche}, {Antoja}, {Arcay}, {Astraatmadja}, {Bach},
  {Baker}, {Balaguer-N{\'u}{\~n}ez}, {Balm}, {Barache}, {Barata}, {Barbato},
  {Barblan}, {Barklem}, {Barrado}, {Barros}, {Barstow}, {Bartholom{\'e}
  Mu{\~n}oz}, {Bassilana}, {Becciani}, {Bellazzini}, {Berihuete}, {Bertone},
  {Bianchi}, {Bienaym{\'e}}, {Blanco-Cuaresma}, {Boch}, {Boeche}, {Bombrun},
  {Borrachero}, {Bossini}, {Bouquillon}, {Bourda}, {Bragaglia}, {Bramante},
  {Breddels}, {Bressan}, {Brouillet}, {Br{\"u}semeister}, {Brugaletta},
  {Bucciarelli}, {Burlacu}, {Busonero}, {Butkevich}, {Buzzi}, {Caffau},
  {Cancelliere}, {Cannizzaro}, {Cantat-Gaudin}, {Carballo}, {Carlucci},
  {Carrasco}, {Casamiquela}, {Castellani}, {Castro-Ginard}, {Charlot},
  {Chemin}, {Chiavassa}, {Cocozza}, {Costigan}, {Cowell}, {Crifo}, {Crosta},
  {Crowley}, {Cuypers}, {Dafonte}, {Damerdji}, {Dapergolas}, {David}, {David},
  {de Laverny}, {De Luise}, {De March}, {de Martino}, {de Souza}, {de Torres},
  {Debosscher}, {del Pozo}, {Delbo}, {Delgado}, {Delgado}, {Di Matteo},
  {Diakite}, {Diener}, {Distefano}, {Dolding}, {Drazinos}, {Dur{\'a}n},
  {Edvardsson}, {Enke}, {Eriksson}, {Esquej}, {Eynard Bontemps}, {Fabre},
  {Fabrizio}, {Faigler}, {Falc{\~a}o}, {Farr{\`a}s Casas}, {Federici},
  {Fedorets}, {Fernique}, {Figueras}, {Filippi}, {Findeisen}, {Fonti},
  {Fraile}, {Fraser}, {Fr{\'e}zouls}, {Gai}, {Galleti}, {Garabato},
  {Garc{\'\i}a-Sedano}, {Garofalo}, {Garralda}, {Gavel}, {Gavras}, {Gerssen},
  {Geyer}, {Giacobbe}, {Gilmore}, {Girona}, {Giuffrida}, {Glass}, {Gomes},
  {Granvik}, {Gueguen}, {Guerrier}, {Guiraud}, {Guti{\'e}rrez-S{\'a}nchez},
  {Haigron}, {Hatzidimitriou}, {Hauser}, {Haywood}, {Heiter}, {Helmi}, {Heu},
  {Hilger}, {Hobbs}, {Hofmann}, {Holland}, {Huckle}, {Hypki}, {Icardi},
  {Jan{\ss}en}, {Jevardat de Fombelle}, {Jonker}, {Juh{\'a}sz}, {Julbe},
  {Karampelas}, {Kewley}, {Klar}, {Kochoska}, {Kohley}, {Kolenberg},
  {Kontizas}, {Kontizas}, {Koposov}, {Kordopatis}, {Kostrzewa-Rutkowska},
  {Koubsky}, {Lambert}, {Lanza}, {Lasne}, {Lavigne}, {Le Fustec}, {Le
  Poncin-Lafitte}, {Lebreton}, {Leccia}, {Leclerc}, {Lecoeur-Taibi},
  {Lenhardt}, {Leroux}, {Liao}, {Licata}, {Lindstr{\o}m}, {Lister}, {Livanou},
  {Lobel}, {L{\'o}pez}, {Managau}, {Mann}, {Mantelet}, {Marchal}, {Marchant},
  {Marconi}, {Marinoni}, {Marschalk{\'o}}, {Marshall}, {Martino}, {Marton},
  {Mary}, {Massari}, {Matijevi{\v{c}}}, {Mazeh}, {McMillan}, {Messina},
  {Michalik}, {Millar}, {Molina}, {Molinaro}, {Moln{\'a}r}, {Montegriffo},
  {Mor}, {Morbidelli}, {Morel}, {Morris}, {Mulone}, {Muraveva}, {Musella},
  {Nelemans}, {Nicastro}, {Noval}, {O'Mullane}, {Ord{\'e}novic},
  {Ord{\'o}{\~n}ez-Blanco}, {Osborne}, {Pagani}, {Pagano}, {Pailler},
  {Palacin}, {Palaversa}, {Panahi}, {Pawlak}, {Piersimoni}, {Pineau}, {Plachy},
  {Plum}, {Poggio}, {Poujoulet}, {Pr{\v{s}}a}, {Pulone}, {Racero}, {Ragaini},
  {Rambaux}, {Ramos-Lerate}, {Regibo}, {Reyl{\'e}}, {Riclet}, {Ripepi}, {Riva},
  {Rivard}, {Rixon}, {Roegiers}, {Roelens}, {Romero-G{\'o}mez}, {Rowell},
  {Royer}, {Ruiz-Dern}, {Sadowski}, {Sagrist{\`a} Sell{\'e}s}, {Sahlmann},
  {Salgado}, {Salguero}, {Sanna}, {Santana-Ros}, {Sarasso}, {Savietto},
  {Schultheis}, {Sciacca}, {Segol}, {Segovia}, {S{\'e}gransan}, {Shih},
  {Siltala}, {Silva}, {Smart}, {Smith}, {Solano}, {Solitro}, {Sordo}, {Soria
  Nieto}, {Souchay}, {Spagna}, {Spoto}, {Stampa}, {Steele},
  {Steidelm{\"u}ller}, {Stephenson}, {Stoev}, {Suess}, {Surdej}, {Szabados},
  {Szegedi-Elek}, {Tapiador}, {Taris}, {Tauran}, {Taylor}, {Teixeira},
  {Terrett}, {Teyssandier}, {Thuillot}, {Titarenko}, {Torra Clotet}, {Turon},
  {Ulla}, {Utrilla}, {Uzzi}, {Vaillant}, {Valentini}, {Valette}, {van Elteren},
  {Van Hemelryck}, {van Leeuwen}, {Vaschetto}, {Vecchiato}, {Veljanoski},
  {Viala}, {Vicente}, {Vogt}, {von Essen}, {Voss}, {Votruba}, {Voutsinas},
  {Walmsley}, {Weiler}, {Wertz}, {Wevers}, {Wyrzykowski}, {Yoldas},
  {{\v{Z}}erjal}, {Ziaeepour}, {Zorec}, {Zschocke}, {Zucker}, {Zurbach}, \&
  {Zwitter}}]{GaiaDR2}
{Gaia Collaboration}, {Brown}, A.~G.~A., {Vallenari}, A., {et~al.} 2018, \aap,
  616, A1

\bibitem[{{Gaia Collaboration} {et~al.}(2021){Gaia Collaboration}, {Smart},
  {Sarro}, {Rybizki}, {Reyl{\'e}}, {Robin}, {Hambly}, {Abbas}, {Barstow}, {de
  Bruijne}, {Bucciarelli}, {Carrasco}, {Cooper}, {Hodgkin}, {Masana},
  {Michalik}, {Sahlmann}, {Sozzetti}, {Brown}, {Vallenari}, {Prusti},
  {Babusiaux}, {Biermann}, {Creevey}, {Evans}, {Eyer}, {Hutton}, {Jansen},
  {Jordi}, {Klioner}, {Lammers}, {Lindegren}, {Luri}, {Mignard}, {Panem},
  {Pourbaix}, {Randich}, {Sartoretti}, {Soubiran}, {Walton}, {Arenou},
  {Bailer-Jones}, {Bastian}, {Cropper}, {Drimmel}, {Katz}, {Lattanzi}, {van
  Leeuwen}, {Bakker}, {Casta{\~n}eda}, {De Angeli}, {Ducourant}, {Fabricius},
  {Fouesneau}, {Fr{\'e}mat}, {Guerra}, {Guerrier}, {Guiraud}, {Jean-Antoine
  Piccolo}, {Messineo}, {Mowlavi}, {Nicolas}, {Nienartowicz}, {Pailler},
  {Panuzzo}, {Riclet}, {Roux}, {Seabroke}, {Sordo}, {Tanga}, {Th{\'e}venin},
  {Gracia-Abril}, {Portell}, {Teyssier}, {Altmann}, {Andrae}, {Bellas-Velidis},
  {Benson}, {Berthier}, {Blomme}, {Brugaletta}, {Burgess}, {Busso}, {Carry},
  {Cellino}, {Cheek}, {Clementini}, {Damerdji}, {Davidson}, {Delchambre},
  {Dell'Oro}, {Fern{\'a}ndez-Hern{\'a}ndez}, {Galluccio}, {Garc{\'\i}a-Lario},
  {Garcia-Reinaldos}, {Gonz{\'a}lez-N{\'u}{\~n}ez}, {Gosset}, {Haigron},
  {Halbwachs}, {Harrison}, {Hatzidimitriou}, {Heiter}, {Hern{\'a}ndez},
  {Hestroffer}, {Holl}, {Jan{\ss}en}, {Jevardat de Fombelle}, {Jordan},
  {Krone-Martins}, {Lanzafame}, {L{\"o}ffler}, {Lorca}, {Manteiga}, {Marchal},
  {Marrese}, {Moitinho}, {Mora}, {Muinonen}, {Osborne}, {Pancino}, {Pauwels},
  {Recio-Blanco}, {Richards}, {Riello}, {Rimoldini}, {Roegiers}, {Siopis},
  {Smith}, {Ulla}, {Utrilla}, {van Leeuwen}, {van Reeven}, {Abreu Aramburu},
  {Accart}, {Aerts}, {Aguado}, {Ajaj}, {Altavilla}, {{\'A}lvarez}, {{\'A}lvarez
  Cid-Fuentes}, {Alves}, {Anderson}, {Anglada Varela}, {Antoja}, {Audard},
  {Baines}, {Baker}, {Balaguer-N{\'u}{\~n}ez}, {Balbinot}, {Balog}, {Barache},
  {Barbato}, {Barros}, {Bartolom{\'e}}, {Bassilana}, {Bauchet},
  {Baudesson-Stella}, {Becciani}, {Bellazzini}, {Bernet}, {Bertone}, {Bianchi},
  {Blanco-Cuaresma}, {Boch}, {Bombrun}, {Bossini}, {Bouquillon}, {Bragaglia},
  {Bramante}, {Breedt}, {Bressan}, {Brouillet}, {Burlacu}, {Busonero},
  {Butkevich}, {Buzzi}, {Caffau}, {Cancelliere}, {C{\'a}novas},
  {Cantat-Gaudin}, {Carballo}, {Carlucci}, {Carnerero}, {Casamiquela},
  {Castellani}, {Castro-Ginard}, {Castro Sampol}, {Chaoul}, {Charlot},
  {Chemin}, {Chiavassa}, {Cioni}, {Comoretto}, {Cornez}, {Cowell}, {Crifo},
  {Crosta}, {Crowley}, {Dafonte}, {Dapergolas}, {David}, {David}, {de Laverny},
  {De Luise}, {De March}, {De Ridder}, {de Souza}, {de Teodoro}, {de Torres},
  {del Peloso}, {del Pozo}, {Delgado}, {Delgado}, {Delisle}, {Di Matteo},
  {Diakite}, {Diener}, {Distefano}, {Dolding}, {Eappachen}, {Edvardsson},
  {Enke}, {Esquej}, {Fabre}, {Fabrizio}, {Faigler}, {Fedorets}, {Fernique},
  {Fienga}, {Figueras}, {Fouron}, {Fragkoudi}, {Fraile}, {Franke}, {Gai},
  {Garabato}, {Garcia-Gutierrez}, {Garc{\'\i}a-Torres}, {Garofalo}, {Gavras},
  {Gerlach}, {Geyer}, {Giacobbe}, {Gilmore}, {Girona}, {Giuffrida}, {Gomel},
  {Gomez}, {Gonzalez-Santamaria}, {Gonz{\'a}lez-Vidal}, {Granvik},
  {Guti{\'e}rrez-S{\'a}nchez}, {Guy}, {Hauser}, {Haywood}, {Helmi}, {Hidalgo},
  {Hilger}, {H{\l}adczuk}, {Hobbs}, {Holland}, {Huckle}, {Jasniewicz},
  {Jonker}, {Juaristi Campillo}, {Julbe}, {Karbevska}, {Kervella}, {Khanna},
  {Kochoska}, {Kontizas}, {Kordopatis}, {Korn}, {Kostrzewa-Rutkowska},
  {Kruszy{\'n}ska}, {Lambert}, {Lanza}, {Lasne}, {Le Campion}, {Le Fustec},
  {Lebreton}, {Lebzelter}, {Leccia}, {Leclerc}, {Lecoeur-Taibi}, {Liao},
  {Licata}, {Lindstr{\o}m}, {Lister}, {Livanou}, {Lobel}, {Madrero Pardo},
  {Managau}, {Mann}, {Marchant}, {Marconi}, {Marcos Santos}, {Marinoni},
  {Marocco}, {Marshall}, {Martin Polo}, {Mart{\'\i}n-Fleitas}, {Masip},
  {Massari}, {Mastrobuono-Battisti}, {Mazeh}, {McMillan}, {Messina}, {Millar},
  {Mints}, {Molina}, {Molinaro}, {Moln{\'a}r}, {Montegriffo}, {Mor},
  {Morbidelli}, {Morel}, {Morris}, {Mulone}, {Munoz}, {Muraveva}, {Murphy},
  {Musella}, {Noval}, {Ord{\'e}novic}, {Orr{\`u}}, {Osinde}, {Pagani},
  {Pagano}, {Palaversa}, {Palicio}, {Panahi}, {Pawlak}, {Pe{\~n}alosa
  Esteller}, {Penttil{\"a}}, {Piersimoni}, {Pineau}, {Plachy}, {Plum},
  {Poggio}, {Poretti}, {Poujoulet}, {Pr{\v{s}}a}, {Pulone}, {Racero},
  {Ragaini}, {Rainer}, {Raiteri}, {Rambaux}, {Ramos}, {Ramos-Lerate}, {Re
  Fiorentin}, {Regibo}, {Ripepi}, {Riva}, {Rixon}, {Robichon}, {Robin},
  {Roelens}, {Rohrbasser}, {Romero-G{\'o}mez}, {Rowell}, {Royer}, {Rybicki},
  {Sadowski}, {Sagrist{\`a} Sell{\'e}s}, {Salgado}, {Salguero}, {Samaras},
  {Sanchez Gimenez}, {Sanna}, {Santove{\~n}a}, {Sarasso}, {Schultheis},
  {Sciacca}, {Segol}, {Segovia}, {S{\'e}gransan}, {Semeux}, {Shahaf},
  {Siddiqui}, {Siebert}, {Siltala}, {Slezak}, {Solano}, {Solitro}, {Souami},
  {Souchay}, {Spagna}, {Spoto}, {Steele}, {Steidelm{\"u}ller}, {Stephenson},
  {S{\"u}veges}, {Szabados}, {Szegedi-Elek}, {Taris}, {Tauran}, {Taylor},
  {Teixeira}, {Thuillot}, {Tonello}, {Torra}, {Torra}, {Turon}, {Unger},
  {Vaillant}, {van Dillen}, {Vanel}, {Vecchiato}, {Viala}, {Vicente},
  {Voutsinas}, {Weiler}, {Wevers}, {Wyrzykowski}, {Yoldas}, {Yvard}, {Zhao},
  {Zorec}, {Zucker}, {Zurbach}, \& {Zwitter}}]{GCNS2021}
{Gaia Collaboration}, {Smart}, R.~L., {Sarro}, L.~M., {et~al.} 2021, \aap, 649,
  A6

\bibitem[{Gaia~Collaboration(2022k)}]{GaiaDR3}
Gaia~Collaboration, Vallenari, A. e.~a. 2022k, Astronomy \& Astrophysics, in
  prep

\bibitem[{Giacalone \& Dressing(2020)}]{giacalone2020triceratops}
Giacalone, S., \& Dressing, C.~D. 2020, Astrophysics Source Code Library, ascl

\bibitem[{Giacalone {et~al.}(2021)Giacalone, Dressing, Jensen, Collins, Ricker,
  Vanderspek, Seager, Winn, Jenkins, Barclay, {et~al.}}]{giacalone2021vetting}
Giacalone, S., Dressing, C.~D., Jensen, E.~L., {et~al.} 2021, The Astronomical
  Journal, 161, 24

\bibitem[{Giacalone {et~al.}(2022)Giacalone, Dressing, Hedges, Kostov, Collins,
  Jensen, Yahalomi, Bieryla, Ciardi, Howell,
  {et~al.}}]{giacalone2022validation}
Giacalone, S., Dressing, C.~D., Hedges, C., {et~al.} 2022, The Astronomical
  Journal, 163, 99

\bibitem[{Ginzburg {et~al.}(2018)Ginzburg, Schlichting, \&
  Sari}]{ginzburg2018core}
Ginzburg, S., Schlichting, H.~E., \& Sari, R. 2018, Monthly Notices of the
  Royal Astronomical Society, 476, 759

\bibitem[{{Gray} {et~al.}(2006){Gray}, {Corbally}, {Garrison}, {McFadden},
  {Bubar}, {McGahee}, {O'Donoghue}, \& {Knox}}]{Gray2006}
{Gray}, R.~O., {Corbally}, C.~J., {Garrison}, R.~F., {et~al.} 2006, \aj, 132,
  161

\bibitem[{Guenther {et~al.}(2017)Guenther, Barrag{\'a}n, Dai, Gandolfi, Hirano,
  Fridlund, Fossati, Chau, Helled, Korth, {et~al.}}]{guenther2017k2}
Guenther, E., Barrag{\'a}n, O., Dai, F., {et~al.} 2017, Astronomy \&
  Astrophysics, 608, A93

\bibitem[{Guerrero {et~al.}(2021)Guerrero, Seager, Huang, Vanderburg, Soto,
  Mireles, Hesse, Fong, Glidden, Shporer, {et~al.}}]{guerrero2021tess}
Guerrero, N.~M., Seager, S., Huang, C.~X., {et~al.} 2021, The Astrophysical
  Journal Supplement Series, 254, 39

\bibitem[{Gupta \& Schlichting(2019)}]{gupta2019sculpting}
Gupta, A., \& Schlichting, H.~E. 2019, Monthly Notices of the Royal
  Astronomical Society, 487, 24

\bibitem[{Gupta \& Schlichting(2020)}]{gupta2020signatures}
---. 2020, Monthly Notices of the Royal Astronomical Society, 493, 792

\bibitem[{H{\'e}brard {et~al.}(2003)H{\'e}brard, {\'E}tangs, Vidal-Madjar,
  D{\'e}sert, \& Ferlet}]{hebrard2003evaporation}
H{\'e}brard, G., {\'E}tangs, A., Vidal-Madjar, A., D{\'e}sert, J.-M., \&
  Ferlet, R. 2003, arXiv preprint astro-ph/0312384

\bibitem[{{Henden} {et~al.}(2016){Henden}, {Templeton}, {Terrell}, {Smith},
  {Levine}, \& {Welch}}]{Henden2016}
{Henden}, A.~A., {Templeton}, M., {Terrell}, D., {et~al.} 2016, VizieR Online
  Data Catalog, II/336

\bibitem[{Holzapfel(2018)}]{holzapfel_coherent_2018}
Holzapfel, W.~B. 2018, Solid State Sciences, 80, 31

\bibitem[{Howell {et~al.}(2016)Howell, Everett, Horch, Winters, Hirsch, Nusdeo,
  \& Scott}]{howell2016speckle}
Howell, S.~B., Everett, M.~E., Horch, E.~P., {et~al.} 2016, The Astrophysical
  Journal Letters, 829, L2

\bibitem[{Howell {et~al.}(2011)Howell, Everett, Sherry, Horch, \&
  Ciardi}]{howell2011speckle}
Howell, S.~B., Everett, M.~E., Sherry, W., Horch, E., \& Ciardi, D.~R. 2011,
  The Astronomical Journal, 142, 19

\bibitem[{Howell {et~al.}(2014)Howell, Sobeck, Haas, Still, Barclay, Mullally,
  Troeltzsch, Aigrain, Bryson, Caldwell, {et~al.}}]{howell2014k2}
Howell, S.~B., Sobeck, C., Haas, M., {et~al.} 2014, Publications of the
  Astronomical Society of the Pacific, 126, 398

\bibitem[{Irwin {et~al.}(2015)Irwin, Berta-Thompson, Charbonneau, Dittmann,
  Falco, Newton, \& Nutzman}]{irwin2015mearth}
Irwin, J.~M., Berta-Thompson, Z.~K., Charbonneau, D., {et~al.} 2015, in 18th
  Cambridge Workshop on Cool Stars, Stellar Systems, and the Sun, Proceedings
  of Lowell Observatory, ed. G.~van Belle \& H.~Harris, 767--772

\bibitem[{Ito \& Ikoma(2021)}]{ito2021hydrodynamic}
Ito, Y., \& Ikoma, M. 2021, Monthly Notices of the Royal Astronomical Society,
  502, 750

\bibitem[{Ito {et~al.}(2015)Ito, Ikoma, Kawahara, Nagahara, Kawashima, \&
  Nakamoto}]{ito2015theoretical}
Ito, Y., Ikoma, M., Kawahara, H., {et~al.} 2015, The Astrophysical Journal,
  801, 144

\bibitem[{Jenkins(2002)}]{jenkins2002impact}
Jenkins, J.~M. 2002, The Astrophysical Journal, 575, 493

\bibitem[{Jenkins {et~al.}(2010)Jenkins, Chandrasekaran, McCauliff, Caldwell,
  Tenenbaum, Li, Klaus, Cote, \& Middour}]{jenkins2010transiting}
Jenkins, J.~M., Chandrasekaran, H., McCauliff, S.~D., {et~al.} 2010, in
  Software and Cyberinfrastructure for Astronomy, Vol. 7740, SPIE, 140--150

\bibitem[{{Jenkins} {et~al.}(2016){Jenkins}, {Twicken}, {McCauliff},
  {Campbell}, {Sanderfer}, {Lung}, {Mansouri-Samani}, {Girouard}, {Tenenbaum},
  {Klaus}, {Smith}, {Caldwell}, {Chacon}, {Henze}, {Heiges}, {Latham},
  {Morgan}, {Swade}, {Rinehart}, \& {Vanderspek}}]{Jenkins2016}
{Jenkins}, J.~M., {Twicken}, J.~D., {McCauliff}, S., {et~al.} 2016, Society of
  Photo-Optical Instrumentation Engineers (SPIE) Conference Series, Vol. 9913,
  {The TESS science processing operations center}, 99133E

\bibitem[{{Jensen}(2013)}]{Jensen:2013}
{Jensen}, E. 2013, {Tapir: A web interface for transit/eclipse observability},
  Astrophysics Source Code Library, ascl:1306.007

\bibitem[{Johansen \& Dorn(2022)}]{johansen2022nucleation}
Johansen, A., \& Dorn, C. 2022, arXiv preprint arXiv:2204.04241

\bibitem[{{Johnson} \& {Apps}(2009)}]{Johnson2009}
{Johnson}, J.~A., \& {Apps}, K. 2009, \apj, 699, 933

\bibitem[{Kempton {et~al.}(2018)Kempton, Bean, Louie, Deming, Koll, Mansfield,
  Christiansen, L{\'o}pez-Morales, Swain, Zellem,
  {et~al.}}]{kempton2018framework}
Kempton, E. M.-R., Bean, J.~L., Louie, D.~R., {et~al.} 2018, Publications of
  the Astronomical Society of the Pacific, 130, 114401

\bibitem[{{Kervella} {et~al.}(2022){Kervella}, {Arenou}, \&
  {Th{\'e}venin}}]{Kervella2022}
{Kervella}, P., {Arenou}, F., \& {Th{\'e}venin}, F. 2022, \aap, 657, A7

\bibitem[{{Kilic} {et~al.}(2017){Kilic}, {Munn}, {Harris}, {von Hippel},
  {Liebert}, {Williams}, {Jeffery}, \& {DeGennaro}}]{Kilic2017}
{Kilic}, M., {Munn}, J.~A., {Harris}, H.~C., {et~al.} 2017, \apj, 837, 162

\bibitem[{Kipping(2013)}]{kipping2013efficient}
Kipping, D.~M. 2013, Monthly Notices of the Royal Astronomical Society, 435,
  2152

\bibitem[{{Kordopatis} {et~al.}(2013){Kordopatis}, {Gilmore}, {Steinmetz},
  {Boeche}, {Seabroke}, {Siebert}, {Zwitter}, {Binney}, {de Laverny},
  {Recio-Blanco}, {Williams}, {Piffl}, {Enke}, {Roeser}, {Bijaoui}, {Wyse},
  {Freeman}, {Munari}, {Carrillo}, {Anguiano}, {Burton}, {Campbell}, {Cass},
  {Fiegert}, {Hartley}, {Parker}, {Reid}, {Ritter}, {Russell}, {Stupar},
  {Watson}, {Bienaym{\'e}}, {Bland-Hawthorn}, {Gerhard}, {Gibson}, {Grebel},
  {Helmi}, {Navarro}, {Conrad}, {Famaey}, {Faure}, {Just}, {Kos},
  {Matijevi{\v{c}}}, {McMillan}, {Minchev}, {Scholz}, {Sharma}, {Siviero}, {de
  Boer}, \& {{\v{Z}}erjal}}]{Kordopatis2013}
{Kordopatis}, G., {Gilmore}, G., {Steinmetz}, M., {et~al.} 2013, \aj, 146, 134

\bibitem[{Kostov {et~al.}(2019)Kostov, Mullally, Quintana, Coughlin, Mullally,
  Barclay, Col{\'o}n, Schlieder, Barentsen, \& Burke}]{kostov2019discovery}
Kostov, V.~B., Mullally, S.~E., Quintana, E.~V., {et~al.} 2019, The
  Astronomical Journal, 157, 124

\bibitem[{Kreidberg(2015)}]{kreidberg2015batman}
Kreidberg, L. 2015, Publications of the Astronomical Society of the Pacific,
  127, 1161

\bibitem[{Kreidberg {et~al.}(2019)Kreidberg, Koll, Morley, Hu, Schaefer,
  Deming, Stevenson, Dittmann, Vanderburg, Berardo,
  {et~al.}}]{kreidberg2019absence}
Kreidberg, L., Koll, D.~D., Morley, C., {et~al.} 2019, Nature, 573, 87

\bibitem[{Kruss \& Wurm(2018)}]{kruss2018seeding}
Kruss, M., \& Wurm, G. 2018, The Astrophysical Journal, 869, 45

\bibitem[{{Kunder} {et~al.}(2017){Kunder}, {Kordopatis}, {Steinmetz},
  {Zwitter}, {McMillan}, {Casagrande}, {Enke}, {Wojno}, {Valentini},
  {Chiappini}, {Matijevi{\v{c}}}, {Siviero}, {de Laverny}, {Recio-Blanco},
  {Bijaoui}, {Wyse}, {Binney}, {Grebel}, {Helmi}, {Jofre}, {Antoja}, {Gilmore},
  {Siebert}, {Famaey}, {Bienaym{\'e}}, {Gibson}, {Freeman}, {Navarro},
  {Munari}, {Seabroke}, {Anguiano}, {{\v{Z}}erjal}, {Minchev}, {Reid},
  {Bland-Hawthorn}, {Kos}, {Sharma}, {Watson}, {Parker}, {Scholz}, {Burton},
  {Cass}, {Hartley}, {Fiegert}, {Stupar}, {Ritter}, {Hawkins}, {Gerhard},
  {Chaplin}, {Davies}, {Elsworth}, {Lund}, {Miglio}, \& {Mosser}}]{Kunder2017}
{Kunder}, A., {Kordopatis}, G., {Steinmetz}, M., {et~al.} 2017, \aj, 153, 75

\bibitem[{{Lallement} {et~al.}(2018){Lallement}, {Capitanio}, {Ruiz-Dern},
  {Danielski}, {Babusiaux}, {Vergely}, {Elyajouri}, {Arenou}, \&
  {Leclerc}}]{Lallement2018}
{Lallement}, R., {Capitanio}, L., {Ruiz-Dern}, L., {et~al.} 2018, \aap, 616,
  A132

\bibitem[{Lam {et~al.}(2021)Lam, Csizmadia, Astudillo-Defru, Bonfils, Gandolfi,
  Padovan, Esposito, Hellier, Hirano, Livingston, {et~al.}}]{lam2021gj}
Lam, K.~W., Csizmadia, S., Astudillo-Defru, N., {et~al.} 2021, Science, 374,
  1271

\bibitem[{Lee \& Chiang(2016)}]{lee2016breeding}
Lee, E.~J., \& Chiang, E. 2016, The Astrophysical Journal, 817, 90

\bibitem[{Lee {et~al.}(2014)Lee, Chiang, \& Ormel}]{lee2014make}
Lee, E.~J., Chiang, E., \& Ormel, C.~W. 2014, The Astrophysical Journal, 797,
  95

\bibitem[{Lee \& Connors(2021)}]{lee2021primordial}
Lee, E.~J., \& Connors, N.~J. 2021, The Astrophysical Journal, 908, 32

\bibitem[{L{\'e}ger {et~al.}(2011)L{\'e}ger, Grasset, Fegley, Codron, Albarede,
  Barge, Barnes, Cance, Carpy, Catalano, {et~al.}}]{leger2011extreme}
L{\'e}ger, A., Grasset, O., Fegley, B., {et~al.} 2011, Icarus, 213, 1

\bibitem[{Li {et~al.}(2019)Li, Tenenbaum, Twicken, Burke, Jenkins, Quintana,
  Rowe, \& Seader}]{li2019kepler}
Li, J., Tenenbaum, P., Twicken, J.~D., {et~al.} 2019, Publications of the
  Astronomical Society of the Pacific, 131, 024506

\bibitem[{Liu {et~al.}(2015)Liu, Hori, Lin, \& Asphaug}]{liu2015giant}
Liu, S.-F., Hori, Y., Lin, D., \& Asphaug, E. 2015, The Astrophysical Journal,
  812, 164

\bibitem[{Lopez {et~al.}(2012)Lopez, Fortney, \& Miller}]{lopez2012thermal}
Lopez, E.~D., Fortney, J.~J., \& Miller, N. 2012, The Astrophysical Journal,
  761, 59

\bibitem[{Lopez \& Rice(2018)}]{lopez2018formation}
Lopez, E.~D., \& Rice, K. 2018, Monthly Notices of the Royal Astronomical
  Society, 479, 5303

\bibitem[{{Luck}(2018)}]{Luck2018}
{Luck}, R.~E. 2018, \aj, 155, 111

\bibitem[{Luque {et~al.}(2021)Luque, Serrano, Molaverdikhani, Nixon,
  Livingston, Guenther, Pall{\'e}, Madhusudhan, Nowak, Korth,
  {et~al.}}]{luque2021planetary}
Luque, R., Serrano, L., Molaverdikhani, K., {et~al.} 2021, Astronomy \&
  Astrophysics, 645, A41

\bibitem[{Malavolta {et~al.}(2018)Malavolta, Mayo, Louden, Rajpaul, Bonomo,
  Buchhave, Kreidberg, Kristiansen, Lopez-Morales, Mortier,
  {et~al.}}]{malavolta2018ultra}
Malavolta, L., Mayo, A.~W., Louden, T., {et~al.} 2018, The Astronomical
  Journal, 155, 107

\bibitem[{{Mamajek} \& {Hillenbrand}(2008)}]{Mamajek2008}
{Mamajek}, E.~E., \& {Hillenbrand}, L.~A. 2008, \apj, 687, 1264

\bibitem[{{Mann} {et~al.}(2015){Mann}, {Feiden}, {Gaidos}, {Boyajian}, \& {von
  Braun}}]{Mann:2015}
{Mann}, A.~W., {Feiden}, G.~A., {Gaidos}, E., {Boyajian}, T., \& {von Braun},
  K. 2015, \apj, 804, 64

\bibitem[{{Mann} {et~al.}(2019){Mann}, {Dupuy}, {Kraus}, {Gaidos}, {Ansdell},
  {Ireland}, {Rizzuto}, {Hung}, {Dittmann}, {Factor}, {Feiden}, {Martinez},
  {Ru{\'\i}z-Rodr{\'\i}guez}, \& {Thao}}]{Mann:2019}
{Mann}, A.~W., {Dupuy}, T., {Kraus}, A.~L., {et~al.} 2019, \apj, 871, 63

\bibitem[{Mansfield {et~al.}(2019)Mansfield, Kite, Hu, Koll, Malik, Bean, \&
  Kempton}]{mansfield2019identifying}
Mansfield, M., Kite, E.~S., Hu, R., {et~al.} 2019, The Astrophysical Journal,
  886, 141

\bibitem[{Marcus {et~al.}(2010)Marcus, Sasselov, Hernquist, \&
  Stewart}]{marcus2010minimum}
Marcus, R.~A., Sasselov, D., Hernquist, L., \& Stewart, S.~T. 2010, The
  Astrophysical Journal Letters, 712, L73

\bibitem[{Martinez {et~al.}(2019)Martinez, Cunha, Ghezzi, \&
  Smith}]{martinez2019spectroscopic}
Martinez, C.~F., Cunha, K., Ghezzi, L., \& Smith, V.~V. 2019, The Astrophysical
  Journal, 875, 29

\bibitem[{{Maxted} {et~al.}(2011){Maxted}, {Anderson}, {Collier Cameron},
  {Hellier}, {Queloz}, {Smalley}, {Street}, {Triaud}, {West}, {Gillon},
  {Lister}, {Pepe}, {Pollacco}, {S{\'e}gransan}, {Smith}, \&
  {Udry}}]{2011PASP..123..547M}
{Maxted}, P.~F.~L., {Anderson}, D.~R., {Collier Cameron}, A., {et~al.} 2011,
  \pasp, 123, 547

\bibitem[{{McCully} {et~al.}(2018){McCully}, {Volgenau}, {Harbeck}, {Lister},
  {Saunders}, {Turner}, {Siiverd}, \& {Bowman}}]{McCully:2018}
{McCully}, C., {Volgenau}, N.~H., {Harbeck}, D.-R., {et~al.} 2018, in Society
  of Photo-Optical Instrumentation Engineers (SPIE) Conference Series, Vol.
  10707, \procspie, 107070K

\bibitem[{{McQuillan} {et~al.}(2014){McQuillan}, {Mazeh}, \&
  {Aigrain}}]{2014ApJS..211...24M}
{McQuillan}, A., {Mazeh}, T., \& {Aigrain}, S. 2014, \apjs, 211, 24

\bibitem[{Ment {et~al.}(2019)Ment, Dittmann, Astudillo-Defru, Charbonneau,
  Irwin, Bonfils, Murgas, Almenara, Forveille, Agol, {et~al.}}]{ment2019second}
Ment, K., Dittmann, J.~A., Astudillo-Defru, N., {et~al.} 2019, The Astronomical
  Journal, 157, 32

\bibitem[{{Mermilliod} {et~al.}(1997){Mermilliod}, {Mermilliod}, \&
  {Hauck}}]{Mermilliod1997}
{Mermilliod}, J.~C., {Mermilliod}, M., \& {Hauck}, B. 1997, \aaps, 124, 349

\bibitem[{Millholland \& Spalding(2020)}]{millholland2020formation}
Millholland, S.~C., \& Spalding, C. 2020, The Astrophysical Journal, 905, 71

\bibitem[{Mocquet {et~al.}(2014)Mocquet, Grasset, \& Sotin}]{mocquet2014very}
Mocquet, A., Grasset, O., \& Sotin, C. 2014, Philosophical Transactions of the
  Royal Society A: Mathematical, Physical and Engineering Sciences, 372,
  20130164

\bibitem[{{Morris} {et~al.}(2017){Morris}, {Twicken}, {Smith}, {Clarke},
  {Jenkins}, {Bryson}, {Girouard}, \& {Klaus}}]{Morris2017}
{Morris}, R.~L., {Twicken}, J.~D., {Smith}, J.~C., {et~al.} 2017, {Kepler Data
  Processing Handbook: Photometric Analysis}, Tech. rep.

\bibitem[{Nardiello {et~al.}(2022)Nardiello, Malavolta, Desidera, Baratella,
  D'Orazi, Messina, Biazzo, Benatti, Damasso, Rajpaul,
  {et~al.}}]{nardiello2022gapstoi1807}
Nardiello, D., Malavolta, L., Desidera, S., {et~al.} 2022, arXiv preprint
  arXiv:2206.03496

\bibitem[{{NASA Exoplanet Archive}(2022)}]{planetarysystemsexoarchive}
{NASA Exoplanet Archive}. 2022, Planetary Systems, doi:10.26133/NEA12

\bibitem[{{NExScI}(2022)}]{https://doi.org/10.26134/exofop5}
{NExScI}. 2022, Exoplanet Follow-up Observing Program Web Service,
  doi:10.26134/EXOFOP5

\bibitem[{{Noyes} {et~al.}(1984){Noyes}, {Hartmann}, {Baliunas}, {Duncan}, \&
  {Vaughan}}]{Noyes1984}
{Noyes}, R.~W., {Hartmann}, L.~W., {Baliunas}, S.~L., {Duncan}, D.~K., \&
  {Vaughan}, A.~H. 1984, \apj, 279, 763

\bibitem[{Nutzman \& Charbonneau(2008)}]{nutzman2008design}
Nutzman, P., \& Charbonneau, D. 2008, Publications of the Astronomical Society
  of the Pacific, 120, 317

\bibitem[{Osborn {et~al.}(2021)Osborn, Armstrong, Cale, Brahm, Wittenmyer, Dai,
  Crossfield, Bryant, Adibekyan, Cloutier, {et~al.}}]{osborn2021toi}
Osborn, A., Armstrong, D.~J., Cale, B., {et~al.} 2021, Monthly Notices of the
  Royal Astronomical Society, 507, 2782

\bibitem[{Owen \& Wu(2017)}]{owen2017evaporation}
Owen, J.~E., \& Wu, Y. 2017, The Astrophysical Journal, 847, 29

\bibitem[{{Paegert} {et~al.}(2021){Paegert}, {Stassun}, {Collins}, {Pepper},
  {Torres}, {Jenkins}, {Twicken}, \& {Latham}}]{Paegert2021}
{Paegert}, M., {Stassun}, K.~G., {Collins}, K.~A., {et~al.} 2021, arXiv
  e-prints, arXiv:2108.04778

\bibitem[{{Pecaut} \& {Mamajek}(2013)}]{Pecaut2013}
{Pecaut}, M.~J., \& {Mamajek}, E.~E. 2013, \apjs, 208, 9

\bibitem[{Pepe {et~al.}(2013)Pepe, Cameron, Latham, Molinari, Udry, Bonomo,
  Buchhave, Charbonneau, Cosentino, Dressing, {et~al.}}]{pepe2013earth}
Pepe, F., Cameron, A.~C., Latham, D.~W., {et~al.} 2013, Nature, 503, 377

\bibitem[{{Pollacco} {et~al.}(2006){Pollacco}, {Skillen}, {Collier Cameron},
  {Christian}, {Hellier}, {Irwin}, {Lister}, {Street}, {West}, {Anderson},
  {Clarkson}, {Deeg}, {Enoch}, {Evans}, {Fitzsimmons}, {Haswell}, {Hodgkin},
  {Horne}, {Kane}, {Keenan}, {Maxted}, {Norton}, {Osborne}, {Parley}, {Ryans},
  {Smalley}, {Wheatley}, \& {Wilson}}]{2006PASP..118.1407P}
{Pollacco}, D.~L., {Skillen}, I., {Collier Cameron}, A., {et~al.} 2006, \pasp,
  118, 1407

\bibitem[{Prusti {et~al.}(2016)Prusti, De~Bruijne, Brown, Vallenari, Babusiaux,
  Bailer-Jones, Bastian, Biermann, Evans, Eyer, {et~al.}}]{prusti2016gaia}
Prusti, T., De~Bruijne, J., Brown, A.~G., {et~al.} 2016, Astronomy \&
  astrophysics, 595, A1

\bibitem[{Pu \& Lai(2019)}]{pu2019low}
Pu, B., \& Lai, D. 2019, Monthly Notices of the Royal Astronomical Society,
  488, 3568

\bibitem[{{Reiners} \& {Zechmeister}(2020)}]{Reiners2020}
{Reiners}, A., \& {Zechmeister}, M. 2020, \apjs, 247, 11

\bibitem[{{Reiners} {et~al.}(2022){Reiners}, {Shulyak}, {K{\"a}pyl{\"a}},
  {Ribas}, {Nagel}, {Zechmeister}, {Caballero}, {Shan}, {Fuhrmeister},
  {Quirrenbach}, {Amado}, {Montes}, {Jeffers}, {Azzaro}, {B{\'e}jar},
  {Chaturvedi}, {Henning}, {K{\"u}rster}, \& {Pall{\'e}}}]{Reiners2022}
{Reiners}, A., {Shulyak}, D., {K{\"a}pyl{\"a}}, P.~J., {et~al.} 2022, arXiv
  e-prints, arXiv:2204.00342

\bibitem[{Ricker {et~al.}(2014)Ricker, Winn, Vanderspek, Latham, Bakos, Bean,
  Berta-Thompson, Brown, Buchhave, Butler, {et~al.}}]{ricker2014transiting}
Ricker, G.~R., Winn, J.~N., Vanderspek, R., {et~al.} 2014, Journal of
  Astronomical Telescopes, Instruments, and Systems, 1, 014003

\bibitem[{Rogers(2015)}]{rogers2015most}
Rogers, L.~A. 2015, The Astrophysical Journal, 801, 41

\bibitem[{Rubie {et~al.}(2015)Rubie, Jacobson, Morbidelli, O’Brien, Young,
  de~Vries, Nimmo, Palme, \& Frost}]{rubie2015accretion}
Rubie, D.~C., Jacobson, S.~A., Morbidelli, A., {et~al.} 2015, Icarus, 248, 89

\bibitem[{Santerne {et~al.}(2018)Santerne, Brugger, Armstrong, Adibekyan,
  Lillo-Box, Gosselin, Aguichine, Almenara, Barrado, Barros,
  {et~al.}}]{santerne2018earth}
Santerne, A., Brugger, B., Armstrong, D., {et~al.} 2018, Nature Astronomy, 2,
  393

\bibitem[{Schaefer \& Fegley(2009)}]{schaefer2009chemistry}
Schaefer, L., \& Fegley, B. 2009, The Astrophysical Journal, 703, L113

\bibitem[{{Schlaufman} \& {Laughlin}(2010)}]{Schlaufman2010}
{Schlaufman}, K.~C., \& {Laughlin}, G. 2010, \aap, 519, A105

\bibitem[{{Schofield} {et~al.}(2019){Schofield}, {Chaplin}, {Huber},
  {Campante}, {Davies}, {Miglio}, {Ball}, {Appourchaux}, {Basu}, {Bedding},
  {Christensen-Dalsgaard}, {Creevey}, {Garc{\'\i}a}, {Handberg}, {Kawaler},
  {Kjeldsen}, {Latham}, {Lund}, {Metcalfe}, {Ricker}, {Serenelli}, {Silva
  Aguirre}, {Stello}, \& {Vanderspek}}]{Schofield2019}
{Schofield}, M., {Chaplin}, W.~J., {Huber}, D., {et~al.} 2019, \apjs, 241, 12

\bibitem[{{Schr{\"o}der} {et~al.}(2009){Schr{\"o}der}, {Reiners}, \&
  {Schmitt}}]{Schroeder2009}
{Schr{\"o}der}, C., {Reiners}, A., \& {Schmitt}, J.~H.~M.~M. 2009, \aap, 493,
  1099

\bibitem[{Schulze {et~al.}(2021)Schulze, Wang, Johnson, Gaudi, Unterborn, \&
  Panero}]{schulze2021probability}
Schulze, J., Wang, J., Johnson, J., {et~al.} 2021, The Planetary Science
  Journal, 2, 113

\bibitem[{Scora {et~al.}(2020)Scora, Valencia, Morbidelli, \&
  Jacobson}]{scora2020chemical}
Scora, J., Valencia, D., Morbidelli, A., \& Jacobson, S. 2020, Monthly Notices
  of the Royal Astronomical Society, 493, 4910

\bibitem[{Scott {et~al.}(2021)Scott, Howell, Gnilka, Stephens, Salinas, Matson,
  Furlan, Horch, Everett, Ciardi, {et~al.}}]{scott2021twin}
Scott, N.~J., Howell, S.~B., Gnilka, C.~L., {et~al.} 2021, Frontiers in
  Astronomy and Space Sciences, 8, 138

\bibitem[{Seager {et~al.}(2007)Seager, Kuchner, Hier-Majumder, \&
  Militzer}]{seager2007mass}
Seager, S., Kuchner, M., Hier-Majumder, C., \& Militzer, B. 2007, The
  Astrophysical Journal, 669, 1279

\bibitem[{Serrano {et~al.}(2022)Serrano, Gandolfi, Mustill, Barrag{\'a}n,
  Korth, Dai, Redfield, Fridlund, Lam, D{\'\i}az, {et~al.}}]{serrano2022low}
Serrano, L.~M., Gandolfi, D., Mustill, A.~J., {et~al.} 2022, Nature Astronomy,
  1

\bibitem[{Silva {et~al.}(2022)Silva, Demangeon, Barros, Armstrong, Otegi,
  Bossini, Mena, Sousa, Adibekyan, Nielsen, {et~al.}}]{silva2022hd}
Silva, T.~A., Demangeon, O., Barros, S., {et~al.} 2022, Astronomy \&
  Astrophysics, 657, A68

\bibitem[{{Smith} {et~al.}(2012){Smith}, {Stumpe}, {Van Cleve}, {Jenkins},
  {Barclay}, {Fanelli}, {Girouard}, {Kolodziejczak}, {McCauliff}, {Morris}, \&
  {Twicken}}]{Smith2012}
{Smith}, J.~C., {Stumpe}, M.~C., {Van Cleve}, J.~E., {et~al.} 2012, \pasp, 124,
  1000

\bibitem[{{Soto} \& {Jenkins}(2018)}]{Soto2018}
{Soto}, M.~G., \& {Jenkins}, J.~S. 2018, \aap, 615, A76

\bibitem[{{Sousa} {et~al.}(2018){Sousa}, {Adibekyan}, {Delgado-Mena}, {Santos},
  {Andreasen}, {Ferreira}, {Tsantaki}, {Barros}, {Demangeon}, {Israelian},
  {Faria}, {Figueira}, {Mortier}, {Brand{\~a}o}, {Montalto}, {Rojas-Ayala}, \&
  {Santerne}}]{Sousa2018}
{Sousa}, S.~G., {Adibekyan}, V., {Delgado-Mena}, E., {et~al.} 2018, \aap, 620,
  A58

\bibitem[{Souto {et~al.}(2022)Souto, Cunha, Smith, Prieto, Covey,
  Garc{\'\i}a-Hern{\'a}ndez, Holtzman, J{\"o}nsson, Mahadevan, Majewski,
  {et~al.}}]{souto2022detailed}
Souto, D., Cunha, K., Smith, V.~V., {et~al.} 2022, The Astrophysical Journal,
  927, 123

\bibitem[{Speagle(2020)}]{speagle2020dynesty}
Speagle, J.~S. 2020, Monthly Notices of the Royal Astronomical Society, 493,
  3132

\bibitem[{{Stassun} {et~al.}(2017){Stassun}, {Collins}, \&
  {Gaudi}}]{Stassun:2017}
{Stassun}, K.~G., {Collins}, K.~A., \& {Gaudi}, B.~S. 2017, \aj, 153, 136

\bibitem[{{Stassun} {et~al.}(2018){Stassun}, {Corsaro}, {Pepper}, \&
  {Gaudi}}]{Stassun:2018}
{Stassun}, K.~G., {Corsaro}, E., {Pepper}, J.~A., \& {Gaudi}, B.~S. 2018, \aj,
  155, 22

\bibitem[{{Stassun} \& {Torres}(2016)}]{Stassun:2016}
{Stassun}, K.~G., \& {Torres}, G. 2016, \aj, 152, 180

\bibitem[{{Stassun} \& {Torres}(2021)}]{StassunTorres:2021}
---. 2021, \apjl, 907, L33

\bibitem[{Stassun {et~al.}(2018)Stassun, Oelkers, Pepper, Paegert, De~Lee,
  Torres, Latham, Charpinet, Dressing, Huber, {et~al.}}]{stassun2018tess}
Stassun, K.~G., Oelkers, R.~J., Pepper, J., {et~al.} 2018, The Astronomical
  Journal, 156, 102

\bibitem[{{Stassun} {et~al.}(2019){Stassun}, {Oelkers}, {Paegert}, {Torres},
  {Pepper}, {De Lee}, {Collins}, {Latham}, {Muirhead}, {Chittidi},
  {Rojas-Ayala}, {Fleming}, {Rose}, {Tenenbaum}, {Ting}, {Kane}, {Barclay},
  {Bean}, {Brassuer}, {Charbonneau}, {Ge}, {Lissauer}, {Mann}, {McLean},
  {Mullally}, {Narita}, {Plavchan}, {Ricker}, {Sasselov}, {Seager}, {Sharma},
  {Shiao}, {Sozzetti}, {Stello}, {Vanderspek}, {Wallace}, \&
  {Winn}}]{Stassun2019}
{Stassun}, K.~G., {Oelkers}, R.~J., {Paegert}, M., {et~al.} 2019, \aj, 158, 138

\bibitem[{{Steinmetz} {et~al.}(2020){Steinmetz}, {Matijevi{\v{c}}}, {Enke},
  {Zwitter}, {Guiglion}, {McMillan}, {Kordopatis}, {Valentini}, {Chiappini},
  {Casagrande}, {Wojno}, {Anguiano}, {Bienaym{\'e}}, {Bijaoui}, {Binney},
  {Burton}, {Cass}, {de Laverny}, {Fiegert}, {Freeman}, {Fulbright}, {Gibson},
  {Gilmore}, {Grebel}, {Helmi}, {Kunder}, {Munari}, {Navarro}, {Parker},
  {Ruchti}, {Recio-Blanco}, {Reid}, {Seabroke}, {Siviero}, {Siebert}, {Stupar},
  {Watson}, {Williams}, {Wyse}, {Anders}, {Antoja}, {Birko}, {Bland-Hawthorn},
  {Bossini}, {Garc{\'\i}a}, {Carrillo}, {Chaplin}, {Elsworth}, {Famaey},
  {Gerhard}, {Jofre}, {Just}, {Mathur}, {Miglio}, {Minchev}, {Monari},
  {Mosser}, {Ritter}, {Rodrigues}, {Scholz}, {Sharma}, {Sysoliatina}, \& {RAVE
  Collaboration}}]{Steinmetz2020}
{Steinmetz}, M., {Matijevi{\v{c}}}, G., {Enke}, H., {et~al.} 2020, \aj, 160, 82

\bibitem[{{Stumpe} {et~al.}(2014){Stumpe}, {Smith}, {Catanzarite}, {Van Cleve},
  {Jenkins}, {Twicken}, \& {Girouard}}]{Stumpe2014}
{Stumpe}, M.~C., {Smith}, J.~C., {Catanzarite}, J.~H., {et~al.} 2014, \pasp,
  126, 100

\bibitem[{{Stumpe} {et~al.}(2012){Stumpe}, {Smith}, {Van Cleve}, {Twicken},
  {Barclay}, {Fanelli}, {Girouard}, {Jenkins}, {Kolodziejczak}, {McCauliff}, \&
  {Morris}}]{Stumpe2012}
{Stumpe}, M.~C., {Smith}, J.~C., {Van Cleve}, J.~E., {et~al.} 2012, \pasp, 124,
  985

\bibitem[{{Tononi} {et~al.}(2019){Tononi}, {Torres}, {Garc{\'\i}a-Berro},
  {Camisassa}, {Althaus}, \& {Rebassa-Mansergas}}]{Tononi2019}
{Tononi}, J., {Torres}, S., {Garc{\'\i}a-Berro}, E., {et~al.} 2019, \aap, 628,
  A52

\bibitem[{Trotta(2008)}]{trotta2008bayes}
Trotta, R. 2008, Contemporary Physics, 49, 71

\bibitem[{{Twicken} {et~al.}(2010){Twicken}, {Clarke}, {Bryson}, {Tenenbaum},
  {Wu}, {Jenkins}, {Girouard}, \& {Klaus}}]{Twicken2010}
{Twicken}, J.~D., {Clarke}, B.~D., {Bryson}, S.~T., {et~al.} 2010, Society of
  Photo-Optical Instrumentation Engineers (SPIE) Conference Series, Vol. 7740,
  {Photometric analysis in the Kepler Science Operations Center pipeline},
  774023

\bibitem[{Twicken {et~al.}(2018)Twicken, Catanzarite, Clarke, Girouard,
  Jenkins, Klaus, Li, McCauliff, Seader, Tenenbaum,
  {et~al.}}]{twicken2018kepler}
Twicken, J.~D., Catanzarite, J.~H., Clarke, B.~D., {et~al.} 2018, Publications
  of the Astronomical Society of the Pacific, 130, 064502

\bibitem[{{Upgren} {et~al.}(1972){Upgren}, {Grossenbacher}, {Penhallow},
  {MacConnell}, \& {Frye}}]{Upgren1972}
{Upgren}, A.~R., {Grossenbacher}, R., {Penhallow}, W.~S., {MacConnell}, D.~J.,
  \& {Frye}, R.~L. 1972, \aj, 77, 486

\bibitem[{Van~Eylen {et~al.}(2018)Van~Eylen, Agentoft, Lundkvist, Kjeldsen,
  Owen, Fulton, Petigura, \& Snellen}]{van2018asteroseismic}
Van~Eylen, V., Agentoft, C., Lundkvist, M., {et~al.} 2018, Monthly Notices of
  the Royal Astronomical Society, 479, 4786

\bibitem[{Veyette {et~al.}(2017)Veyette, Muirhead, Mann, Brewer, Allard, \&
  Homeier}]{veyette2017physically}
Veyette, M.~J., Muirhead, P.~S., Mann, A.~W., {et~al.} 2017, The Astrophysical
  Journal, 851, 26

\bibitem[{Weidenschilling(1978)}]{weidenschilling1978iron}
Weidenschilling, S. 1978, Icarus, 35, 99

\bibitem[{Yee {et~al.}(2017)Yee, Petigura, \& {von Braun}}]{Yee2017}
Yee, S.~W., Petigura, E.~A., \& {von Braun}, K. 2017, The Astrophysical
  Journal, 836, 77

\bibitem[{Zeng {et~al.}(2017)Zeng, Jacobsen, \& Sasselov}]{zeng2017gap}
Zeng, L., Jacobsen, S.~B., \& Sasselov, D.~D. 2017, Research Notes of the
  {AAS}, 1, 32

\bibitem[{Zeng {et~al.}(2021)Zeng, Jacobsen, Hyung, Levi, Nava, Kirk, Piaulet,
  Lacedelli, Sasselov, Petaev, Stewart, Alam, López-Morales, Damasso, \&
  Latham}]{zeng_new_2021}
Zeng, L., Jacobsen, S.~B., Hyung, E., {et~al.} 2021, The Astrophysical Journal,
  923, 247

\end{thebibliography}


\clearpage
\appendix
\renewcommand{\thefigure}{A\arabic{figure}}
\renewcommand{\thetable}{A\arabic{table}}
\setcounter{figure}{0}
\setcounter{table}{0}

\begin{figure*}[htb!]
    \centering
    \includegraphics[width=\textwidth,keepaspectratio]{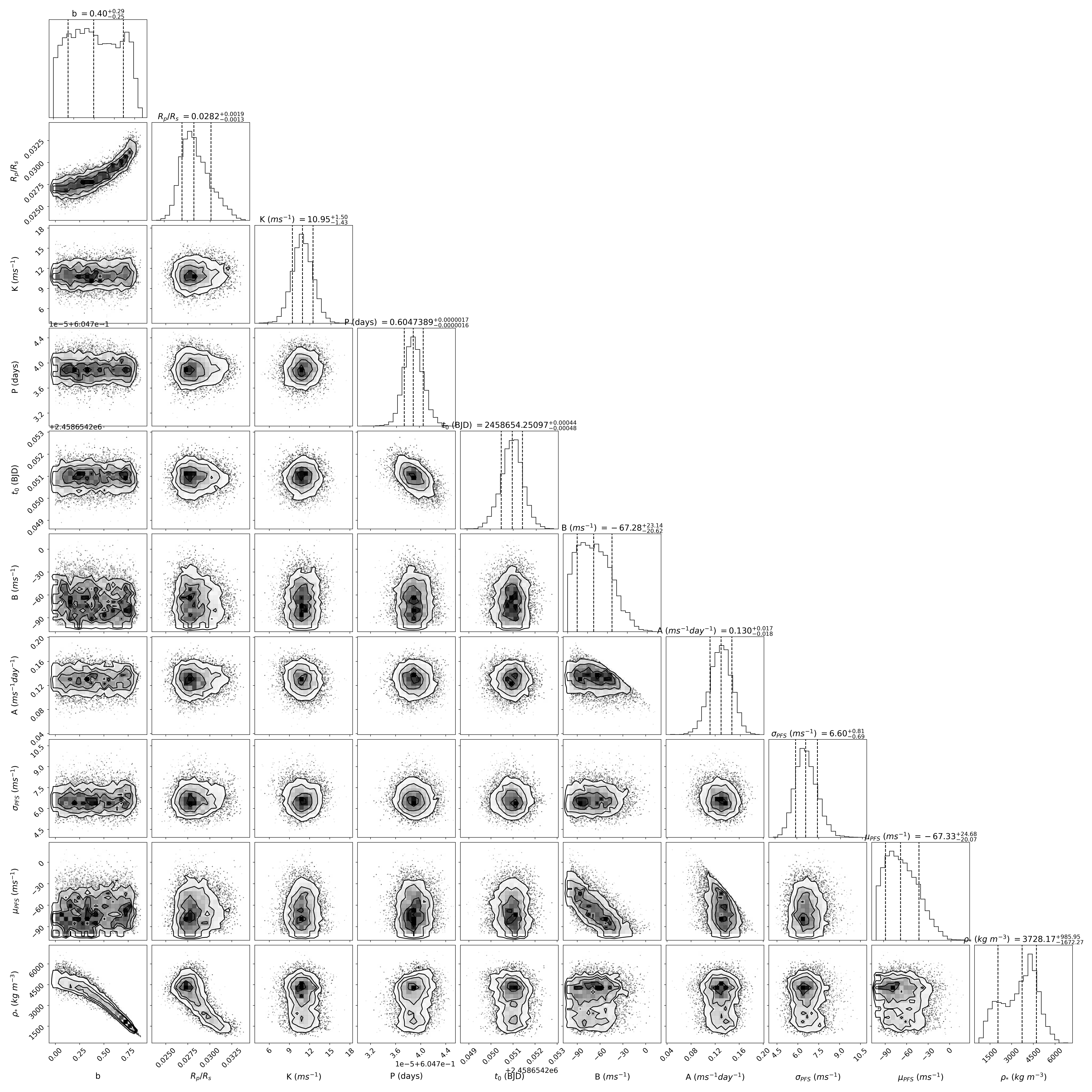}
\caption{Posterior distribution for the joint (photometric + radial velocity) model parameters derived with \texttt{juliet}.  \label{fig:cornerposteriors}}
\end{figure*}

\startlongtable
\begin{deluxetable*}{llccc}
\tabletypesize{\small}
\tablewidth{\linewidth}
\tablecaption{Priors used in our joint analysis of the TOI-1075 system with \texttt{juliet}. The prior labels of $\mathcal{N}$, $\mathcal{U}$, and $\mathcal{LU}$ represent normal, uniform, and log-uniform distributions, respectively, where $\mathcal{N}(\mu, \sigma^2)$ is a normal distribution of the mean $\mu$ and variance $\sigma^2$, and $\mathcal{U}(a, b)$ and
$\mathcal{LU}(a, b)$ are uniform and log-uniform distributions between a and b.}
\tablehead{\colhead{Parameter} & \colhead{Prior} & \colhead{Units and Description}}
\startdata
\smallskip\\\multicolumn{2}{l}{Stellar Parameters:}&\smallskip\\
$\rho_*$ & ~~~~~~~~~~~~~~~~~~~~~~~~$\mathcal{LU}$($10^2, 10^5)$ & Stellar density of TOI-1075 (kg m$^{-3})$\\
\smallskip\\\multicolumn{2}{l}{Planet Parameters:}& TOI-1075 b\smallskip\\
$P_{b}$ & ~~~~~~~~~~~~~~~~~~~~~~~~$\mathcal{N}$($0.6047, 0.0100)$ & Planet period (days)\\
$T_{0,b}$ & ~~~~~~~~~~~~~~~~~~~~~~~~$\mathcal{N}$($2458654.2500, 0.0100)$ & Time of transit center (days)\\
$r_{1,b}$ & ~~~~~~~~~~~~~~~~~~~~~~~~$\mathcal{U}$($0, 1)$ & Parametrization of \citet{espinoza2018efficient} for $p$ and $b$\\
$r_{2,b}$ & ~~~~~~~~~~~~~~~~~~~~~~~~$\mathcal{U}$($0, 1)$ & Parametrization of \citet{espinoza2018efficient} for$p$ and $b$\\
$K_b$ & ~~~~~~~~~~~~~~~~~~~~~~~~$\mathcal{U}$($0, 100)$ & RV semi-amplitude (\ms)\\
$e_b$ & ~~~~~~~~~~~~~~~~~~~~~~~~0.0 (fixed) & Orbital eccentricity\\
$\omega_b$ & ~~~~~~~~~~~~~~~~~~~~~~~~90.0 (fixed) & Periastron angle (deg)\\
\smallskip\\\multicolumn{2}{l}{Photometry Parameters:}\\
$D_{TESS}$ & ~~~~~~~~~~~~~~~~~~~~~~~~1.0 (fixed) & Dilution factor for \TESS\\
$M_{TESS}$ & ~~~~~~~~~~~~~~~~~~~~~~~~$\mathcal{N}$($0, 0.1)$ & Relative flux offset for \TESS\\
$\sigma_{TESS}$ & ~~~~~~~~~~~~~~~~~~~~~~~~$\mathcal{LU}$($0, 5000)$ & Jitter term for \TESS\ light curve (ppm)\\
$q_{1,TESS}$ & ~~~~~~~~~~~~~~~~~~~~~~~~$\mathcal{U}$($0, 1)$ & Quadratic limb-darkening parametrization \citep{kipping2013efficient} \\
$q_{2,TESS}$ & ~~~~~~~~~~~~~~~~~~~~~~~~$\mathcal{U}$($0, 1)$ & Quadratic limb-darkening parametrization \citep{kipping2013efficient}\\
$D_{LCOGT}$ & ~~~~~~~~~~~~~~~~~~~~~~~~1.0 (fixed) & Dilution factor for LCOGT\\
$M_{LCOGT}$ & ~~~~~~~~~~~~~~~~~~~~~~~~$\mathcal{N}$($0, 0.1)$ & Relative flux offset for LCOGT\\
$\sigma_{LCOGT}$ & ~~~~~~~~~~~~~~~~~~~~~~~~$\mathcal{LU}$($0.1, 10^5)$ & Jitter term for LCOGT light curve (ppm)\\
$q_{1,LCOGT}$ & ~~~~~~~~~~~~~~~~~~~~~~~~$\mathcal{U}$($0, 1)$ & Quadratic limb-darkening parametrization \citep{kipping2013efficient} \\
$q_{2,LCOGT}$ & ~~~~~~~~~~~~~~~~~~~~~~~~$\mathcal{U}$($0, 1)$ & Quadratic limb-darkening parametrization \citep{kipping2013efficient}\\
$D_{MEarth}$ & ~~~~~~~~~~~~~~~~~~~~~~~~1.0 (fixed) & Dilution factor for MEarth-South\\
$M_{MEarth}$ & ~~~~~~~~~~~~~~~~~~~~~~~~$\mathcal{N}$($0, 0.1)$ & Relative flux offset for MEarth-South\\
$\sigma_{MEarth}$ & ~~~~~~~~~~~~~~~~~~~~~~~~$\mathcal{LU}$($0.1, 10^5)$ & Jitter term for MEarth-South light curve (ppm)\\
$q_{1,MEarth}$ & ~~~~~~~~~~~~~~~~~~~~~~~~$\mathcal{U}$($0, 1)$ & Quadratic limb-darkening parametrization \citep{kipping2013efficient} \\
$q_{2,MEarth}$ & ~~~~~~~~~~~~~~~~~~~~~~~~$\mathcal{U}$($0, 1)$ & Quadratic limb-darkening parametrization \citep{kipping2013efficient}\\
\smallskip\\\multicolumn{2}{l}{RV Parameters:}\smallskip\\
$\mu_{PFS}$ & ~~~~~~~~~~~~~~~~~~~~~~~~$\mathcal{U}$($-100, 100)$ & Systemic velocity for PFS (\ms) \\
$\sigma_{PFS}$ & ~~~~~~~~~~~~~~~~~~~~~~~~$\mathcal{LU}$($10^{-3}, 100)$ & Jitter term for PFS (\ms) \\
$A$ & ~~~~~~~~~~~~~~~~~~~~~~~~$\mathcal{U}$($-100, 100)$ & Slope of linear long-term RV trend (\ms~day$^{-1}$) \\
$B$ & ~~~~~~~~~~~~~~~~~~~~~~~~$\mathcal{U}$($-100, 100)$ & Intercept of linear long-term RV trend (\ms) \\
\label{tab:julietpriors} \enddata
\end{deluxetable*}


\end{document}